\documentclass[twocolumn]{revtex4}
\usepackage{graphicx}
\usepackage{amssymb}
\usepackage{cancel}
\usepackage{latexsym}
\usepackage{epsfig}
\usepackage{amsmath}
\usepackage{color}
\usepackage[colorlinks]{hyperref}

\hypersetup{
     breaklinks=true,
    pdfstartview={FitH},    
    colorlinks=true,       
    linkcolor=blue,          
    citecolor=red,        
    filecolor=magenta,      
    urlcolor=blue,           
    anchorcolor=green,      
    linktocpage=true
}
\begin{document}

\title{Holographic dark energy in Rastall theory}
\author{S. Ghaffari$^1$\footnote{sh.ghaffari@riaam.ac.ir}, A. A. Mamon$^2$\footnote{abdulla.physics@gmail.com}, H. Moradpour$^1$\footnote{hn.moradpour@maragheh.ac.ir}, A. H. Ziaie$^1$\footnote{ ah.ziaie@maragheh.ac.ir}}
\address{$^1$ Research Institute for Astronomy and Astrophysics of Maragha (RIAAM),
University of Maragheh, P.O. Box 55136-553, Maragheh, Iran\\ $^2$ Department of Physics, Vivekananda Satavarshiki Mahavidyalaya (affiliated
to the Vidyasagar University), Manikpara-721513, West Bengal, India}

\begin{abstract}
Bearing holographic dark energy hypothesis in mind, the ability of
vacuum energy in describing the current accelerated universe is
studied in the framework of Rastall theory. Here, in addition to
the ordinary approach in which it is expected that this energy plays
the role of dark energy, we also address a new approach where the
sum of this energy and Rastall term is responsible for the current
accelerated universe. We also investigate the cosmological outcomes
of using Tsallis entropy in quantifying the energy of fields in
vacuum for both mentioned approaches. The implications of
considering an interaction between the various segments of cosmic
fluid have been addressed in each studied cases. The normalized Hubble parameter for the models have also been plotted and compared that with the $H(z)$ data consisting of 41 data points in the redshift range of $0.07 \leq z \leq 2.36$.
\end{abstract}

\maketitle

\section{Introduction}

Motivated by the long-range nature of gravity, the use of
generalized entropies to study Cosmos has recently been proposed
\cite{non13,EPJC,non20,nun}. In this regard, it has been shown
that the Tsallis entropy can make a bridge between Verlinde and
Padmanabhan hypothesis, and additionally, be combined with
thermodynamic laws to modify Friedmann equations in a way that the
outcomes can explain the accelerated universe without considering
a mysterious dark energy source \cite{non13}. In this regard,
relying on the holographic dark energy hypothesis
\cite{HDE,HDE01,HDE1,HDE2,HDE3,stab,selfint}, and using Tsallis entropy, a
new holographic dark energy, namely Tsallis holographic dark
energy model (THDE), is also introduced in Ref.~\cite{Tavayef}.

In the Einstein framework, where the horizon entropy meets the
Bekenstein entropy, while there is not any mutual interaction
between the Cosmos sectors, both HDE models, obtained by taking
into account the radius of apparent horizon as the IR cutoff, and
pressureless source are scaled with the same function of Hubble
parameter \cite{HDE3}. This difficulty may be solved by using
generalized entropy based on HDE models such as THDE \cite{Tavayef}.
There are observational and theoretical works admitting the
breakdown of the conservation law (the backbone of general
relativity) in curved spacetime \cite{prl,mahep,non20,corda},
a point which has firstly been noted by P. Rastall \cite{rastall}.

Although, Rastall gravity produces acceptable and suitable
predictions and explanations for various gravitational and
cosmological phenomena \cite{corda,ahep,mnras,plb1c,mpla1}, it is
shown that a dark energy-like source is still needed to describe
the current accelerated universe in this framework, if we assume that the universe is filled by cosmic fluids with constant equation of state
\cite{darkrastall,darkrastall1}. A recent study addresses its conflicts and agreements with observations by considering a universe in which a constant plays the role of vacuum energy (as the dark energy candidate) and other cosmic fluids have constant equation of state (Rastall-$\Lambda$CDM model) \cite{aks}. Their study also focus on linear perturbations, while it seems that non-linear perturbations reveal differences between the Rastall cosmology and the $\Lambda$CDM model \cite{darkrastall1}, and it finally signals us that $i$) more observations may be needed to decide about allowed values of the parameter of Rastall theory, and $ii$) a general form of this theory may solve confusion in the current value of Hubble parameter \cite{aks}.

In fact, using the holographic hypothesis, one can get an estimation
for the vacuum energy helping us in studying its cosmological
consequences \cite{HDE}. In the next section, we are going to study
the cosmological implications of vacuum energy in the Rastall
framework by employing the holographic hypothesis, and using horizon
entropy obtained by applying the thermodynamic laws to horizon in
Rastall theory \cite{ahep,plb1c}. In Sec.~($\textmd{III}$), applying
Tsallis entropy to horizon, we shall investigate the ability of
modelling dark energy by using THDE in the Rastall framework. In
each cases, the effects of considering a mutual interaction between
dark energy candidate and other segments of cosmos are also studied.
Moreover, we also propose a new candidate for dark energy i.e., the
sum of vacuum energy and Rastall term. The ability of considering
this hypothesis in describing the current Cosmos has also been
studied in Secs.~($\textmd{II}$) and~($\textmd{III}$). The observational constraints are also addressed throughout the paper. The last
section includes the summary of work, and the units have been set so that $c=\hbar=k_B=1$.

\section{HDE in Rastall theory}

Since the WMAP data indicates a flat universe \cite{roos}, here, we
only focus on flat FRW universe with line element

\begin{eqnarray}\label{rw}
ds^2=-dt^2+a(t)^2[dr^2+r^2(d\theta^2+\sin(\theta)^2d\phi^2)],
\end{eqnarray}

\noindent in which $a(t)$ is called the scale factor. The apparent
horizon of this spacetime, as a proper casual boundary, is also
located at \cite{ahep,plb1c,non20}

\begin{eqnarray}\label{ah}
\tilde{r}_A=\frac{1}{H},
\end{eqnarray}

\noindent and hence, $A=\frac{4\pi}{H^2}$ is the horizon area. The
Rastall field equations are written as \cite{rastall}

\begin{eqnarray}\label{r1}
G_{\mu \nu}+\kappa\lambda g_{\mu \nu}R=\kappa T_{\mu \nu},
\end{eqnarray}

\noindent leading to

\begin{equation}\label{friedman1}
(12\kappa\lambda-3)H^2+6\kappa\lambda\dot{H}=-\kappa\rho,
\end{equation}

\noindent and

\begin{equation}\label{friedman2}
(12\kappa\lambda-3)H^2+(6\kappa\lambda-2)\dot{H}=\kappa p,
\end{equation}

\noindent as the corresponding Friedmann equations. In the above
equations, $\lambda$ and $k$ denote the Rastall parameter and the
Rastall gravitational coupling, respectively. Considering the
Newtonian limit, one can obtain \cite{rastall,ahep}

\begin{eqnarray}\label{eta}
\kappa=\frac{4\eta-1}{6\eta-1}\kappa_G,\ \
\lambda=\frac{\eta(6\eta-1)}{(4\eta-1)\kappa_G},
\end{eqnarray}

\noindent in which $\kappa_G=8\pi G$ is the Einstein gravitational
coupling and $\eta=\kappa\lambda$. Continuity equation governing
the cosmic fluid with energy density $\rho$ and pressure $p$ is
\cite{plb1c,non20}

\begin{equation}\label{cont}
(\frac{3\eta-1}{4\eta-1})\dot{\rho}+(\frac{3\eta}{4\eta-1})\dot{p}+3H(\rho+p)=0.
\end{equation}

\noindent If the Cosmos is filled by a pressureless source with energy
density $\rho_m$ and another fluid with energy density
$\rho_\Lambda$ and pressure $p_\Lambda$, then the Friedmann
equations take the form

\begin{eqnarray}\label{friedman3}
&&(12\eta-3)H^2+6\eta\dot{H}=\frac{4\eta-1}{1-6\eta}\kappa_G(\rho_{\Lambda}+\rho_m),\nonumber\\
&&(12\eta-3)H^2+(6\eta-2)\dot{H}=\frac{4\eta-1}{6\eta-1}\kappa_Gp_\Lambda,
\end{eqnarray}

\noindent leading to

\begin{eqnarray}\label{Rey}
\dot{H}=\frac{4\eta-1}{2(1-6\eta)}\kappa_G(\rho_\Lambda+\rho_m+p_\Lambda).
\end{eqnarray}

\noindent Whenever there is not any mutual interaction between 
both dark components. Eq.~(\ref{cont}) is decomposed to

\begin{equation}\label{cont1}
(\frac{3\eta-1}{4\eta-1})\dot{\rho}_m+3H\rho_m=0\rightarrow\rho_m=\rho_0a^{\frac{3(1-4\eta)}{3\eta-1}},
\end{equation}

\noindent where $\rho_0$ is the integration constant, and

\begin{equation}\label{cont2}
(\frac{3\eta-1}{4\eta-1})\dot{\rho}_\Lambda+(\frac{3\eta}{4\eta-1})\dot{p}_\Lambda+3H(\rho_\Lambda+p_\Lambda)=0.
\end{equation}

Since, in the Rastall framework, we have
$S=(\frac{6\eta-1}{4\eta-1})\frac{A}{4G}$ for the horizon entropy
\cite{plb1c,ahep}, by following the original HDE hypothesis
\cite{HDE}, one easily reaches

\begin{eqnarray}\label{hder}
\rho_\Lambda=\frac{3B}{8\pi G}(\frac{6\eta-1}{4\eta-1})H^2,
\end{eqnarray}

\noindent where $B$ is a numerical constant as usual. Inserting
this result in the first line of Eq.~(\ref{friedman3}), we can
check that $\rho_m\approx H^2$ whenever $\eta=0$ or even whenever
$\eta\neq0$ and
$w_\Lambda\equiv\frac{p_\Lambda}{\rho_\Lambda}=constant$.
Therefore, the same as the Einstein framework ($\eta=0$)
\cite{HDE3}, dark energy and $\rho_m$ will be scaled by the same
function of $H$ in the Rastall framework if $\eta\neq0$ and
$w_\Lambda=constant$. It means that we should have $w_\Lambda\neq
constant$ to avoid this difficulty in the Rastall framework.

\subsection{Common approach}

From Eq.~(\ref{friedman3}), by defining critical density $\rho_c$
as $\rho_c\equiv\frac{3H^2}{8\pi G}$, we get

\begin{eqnarray}\label{friedman4}
&&1=\frac{4\eta-1}{6\eta-1}(\Omega_m+\Omega_\Lambda)+\Omega_\eta+4\eta,
\end{eqnarray}

\noindent in which

\begin{eqnarray}\label{omega}
&&\Omega_m=\frac{\rho_m}{\rho_c},\ \rho_\eta\equiv\frac{6\eta\dot{H}}{8\pi G},\\
&&\Omega_\eta=\frac{\rho_\eta}{\rho_c}=\frac{2\eta\dot{H}}{H^2},\
\Omega_\Lambda=\frac{\rho_\Lambda}{\rho_c}=B\frac{6\eta-1}{4\eta-1}.\nonumber
\end{eqnarray}

\noindent Now, using Eqs. (\ref{friedman3}), (\ref{omega}) and
(\ref{hder}), one can obtain the EoS and deceleration parameters
as

\begin{equation}\label{w1}
\omega_\Lambda=\frac{1}{B}\Big(\frac{\Omega_\eta(3\eta-1)}{3\eta}+4\eta-1\Big),
\end{equation}

\noindent and

\begin{equation}\label{q1}
q=-1-\frac{\dot{H}}{H^2}=-1-\frac{\Omega_\eta}{2\eta},
\end{equation}

\noindent respectively, which are, in effect, the same for both
non-interacting and interacting cases. At the classical level, the
stability of an energy source with energy density $\rho$ and
pressure $p$ is also determined by the sign of the sound speed
square ($v_s^2$) evaluated as

\begin{equation}\label{v_s}
v_s^2=\frac{dp_\Lambda}{d\rho_\Lambda}=\frac{\dot{p}_\Lambda}{\dot{\rho}_\Lambda}=w_\Lambda+\frac{\dot{w}_\Lambda\rho_\Lambda}{\dot{\rho}_\Lambda}.
\end{equation}

\noindent In the following, we study the HDE model in the Rastall
theory for both non-interacting and interacting cases .

\begin{figure}[!]
\begin{center}
    \includegraphics[width=8cm]{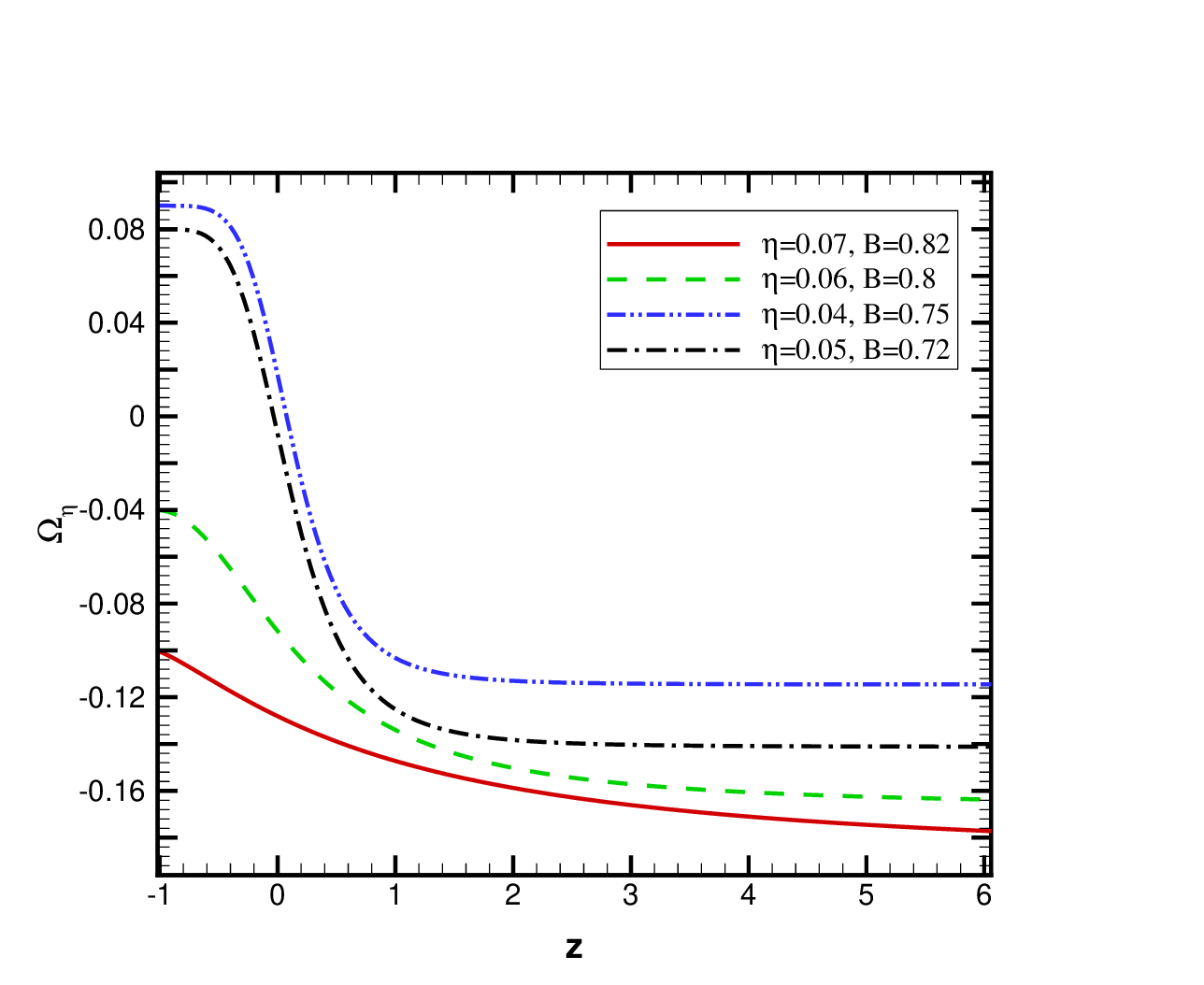}
    \caption{ $\Omega_\eta$ versus $z$ for for non-interacting HDE in Rastall theory
    for some values of $\eta$ and $ B $.}\label{Omega_e}
\end{center}
\end{figure}
\begin{figure}[htp]
\begin{center}
    \includegraphics[width=8cm]{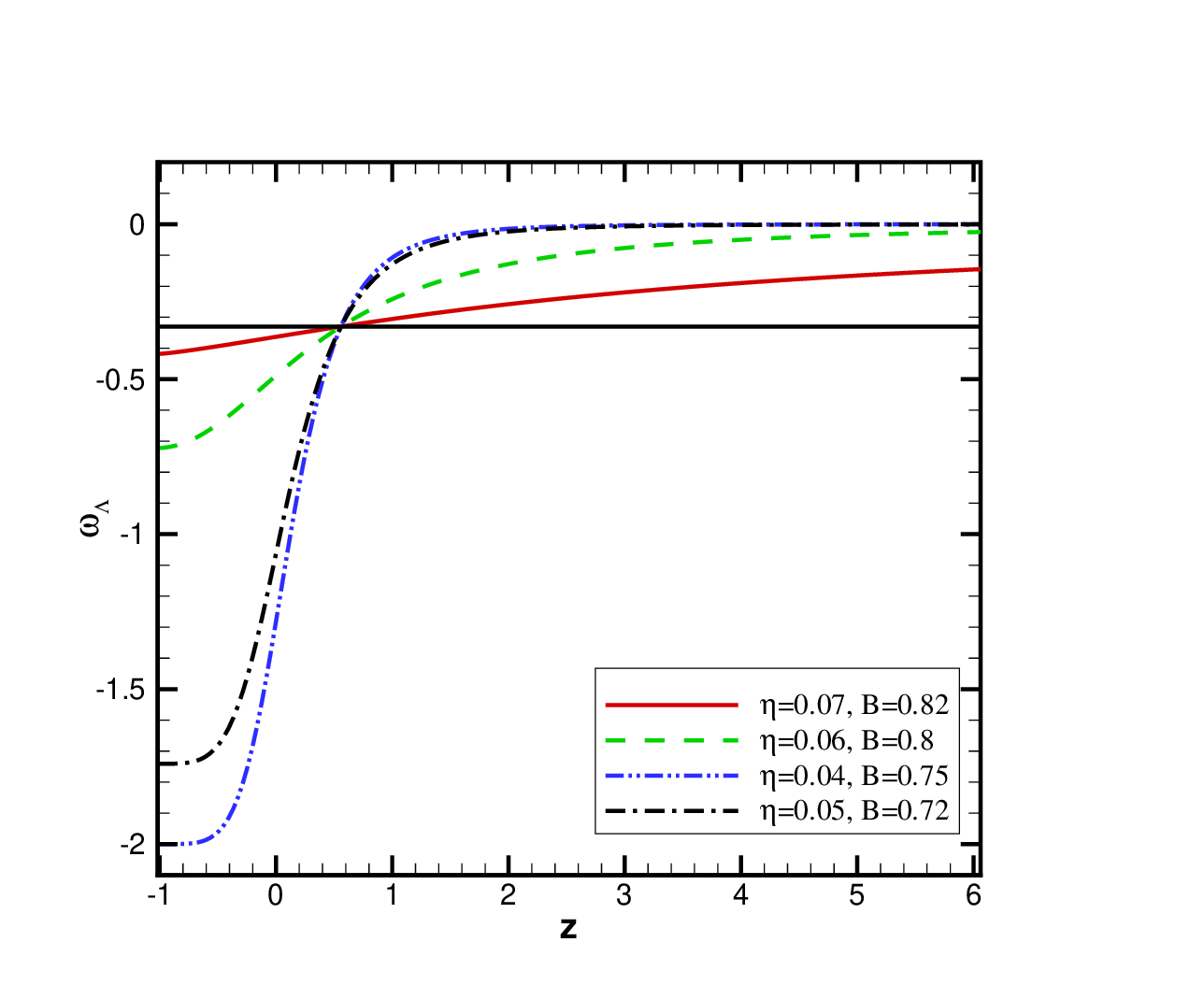}
    \includegraphics[width=7.5cm]{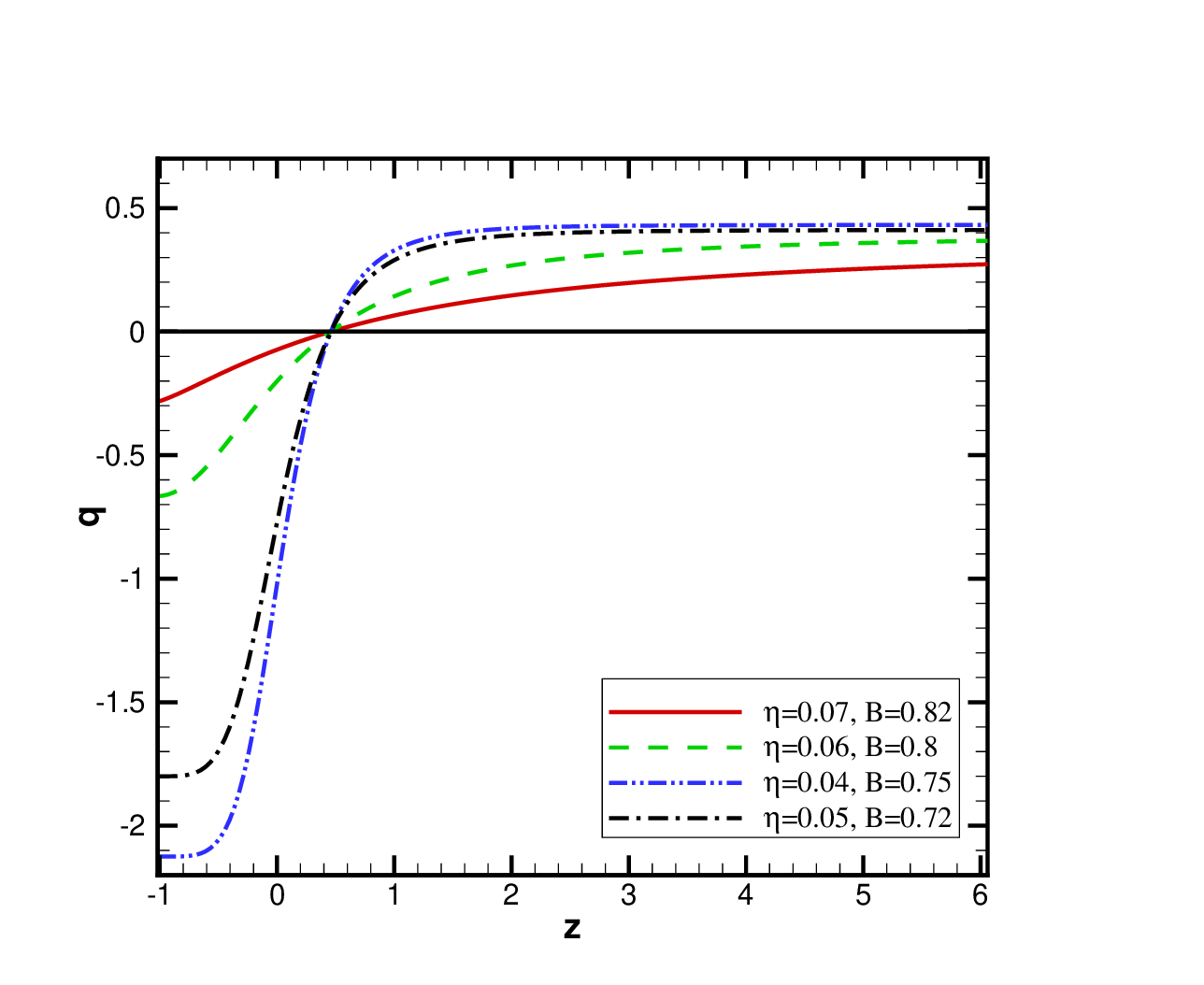}
    \caption{The evolution of the equation of state and deceleration parameters
    with respect to the redshift for non-interacting HDE in Rastall theory.}\label{figw1}
\end{center}
\end{figure}
\begin{figure}[!]
\begin{center}
    \includegraphics[width=8cm]{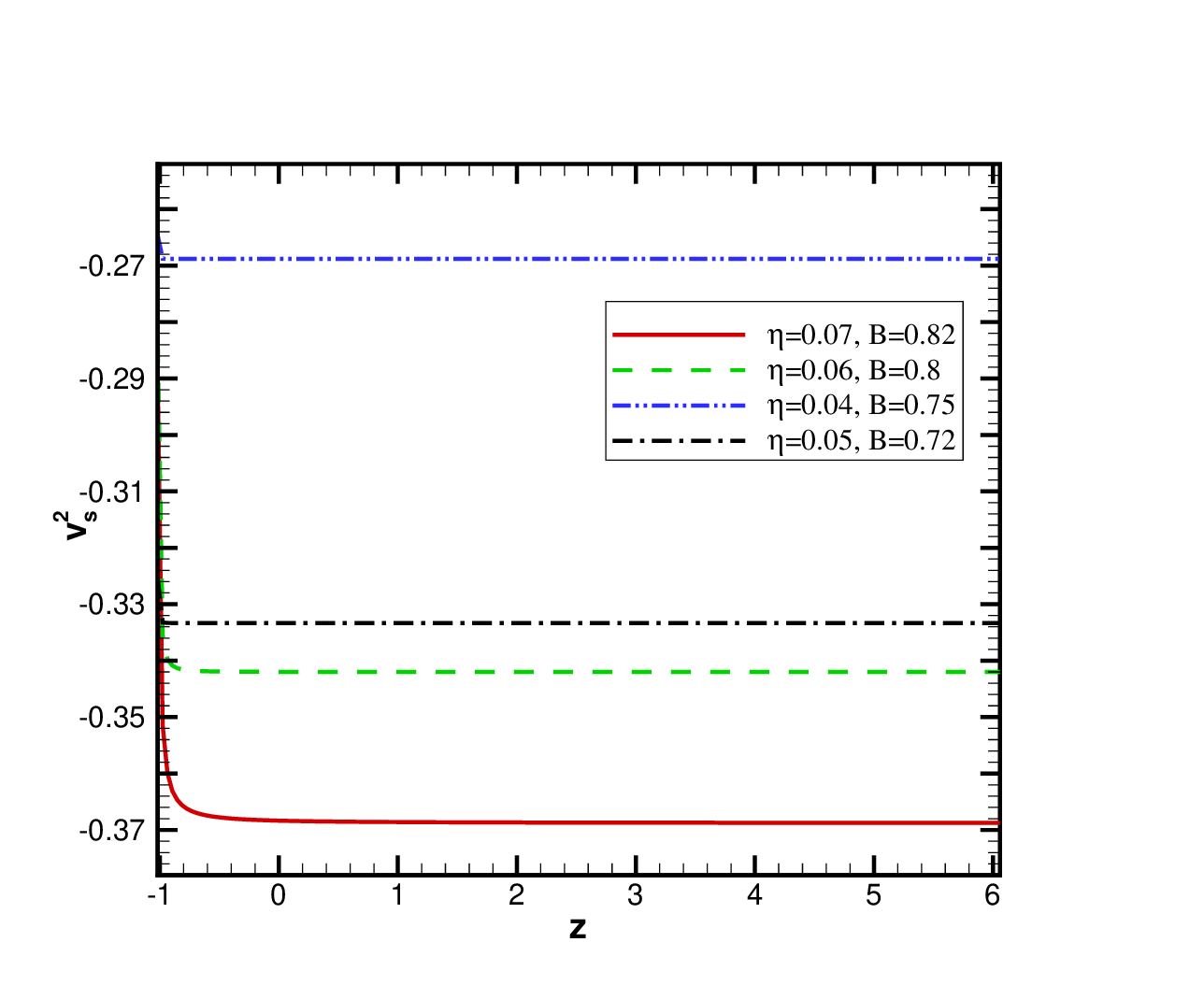}
    \caption{$v_s^2$ versus $z$ for non-interacting HDE in Rastall theory
    for some values of $\eta$ and $ B $.}\label{figv1}
\end{center}
\end{figure}

\subsection*{Non-interacting Case}

Taking the time derivative of the first Friedmann Eq.~(\ref{friedman3}), and combining the result
with Eqs. (\ref{cont1}), (\ref{cont2}), (\ref{hder}) and (\ref{omega}), one finds
\begin{eqnarray}
\frac{\ddot{H}}{H^3}&&=\frac{4\eta-1}{2\eta(1-6\eta)}
\Big[\frac{3\Omega_\eta(6\eta-1)}{3\eta-1}+\\
&&(\Omega_\Lambda+6\eta-1)\Big(\frac{\Omega_\eta}{\eta}+\frac{3(4\eta-1)}{3\eta-1}\Big)\Big],\nonumber
\end{eqnarray}
which can be used to write
\begin{eqnarray}\label{dOmega1}
\Omega_\eta^\prime&=&\frac{\dot{\Omega}_\eta}{H}=\frac{4\eta-1}{1-6\eta}
\Big[\frac{3\Omega_\eta(6\eta-1)}{3\eta-1}+\\
&&(\Omega_\Lambda+6\eta-1)\Big(\frac{\Omega_\eta}{\eta}+\frac{3(4\eta-1)}{3\eta-1}\Big)\Big]-\frac{\Omega_\eta^2}{\eta},\nonumber
\end{eqnarray}
where prime denotes derivative with respect to $ x=\ln a $, and we
used $ \dot{\Omega}_\eta=H\Omega_\eta^\prime $. We have plotted
the evolution of $ \Omega_\eta$  versus redshift parameter $ z $
in Fig. \ref{Omega_e} for some values of $ \eta $ and $ B $. The
behavior of $ \omega_\Lambda $ and $ q $ are also plotted in Fig.
\ref{figw1}. As it is obvious, Universe has a transition from a
deceleration phase to the current accelerated phase at the
redshift $ z\approx 0\cdot6 $ in agreement with the recent
observation data \cite{obser}, and the EoS parameter can not cross
the phantom line. For studying the classical stability of the dark
energy model ($\rho_\Lambda$), by taking the time derivative of
Eq.~(\ref{w1}) and employing Eqs.~(\ref{hder}) and
(\ref{dOmega1}), we reach
\begin{eqnarray}\label{v_s1}
v_s^2&=&\frac{4\eta-1}{B}+\frac{4\eta-1}{3B\eta(1-6\eta)\Omega_\eta}\Big(3\eta(6\eta-1)\Omega_\eta+\nonumber\\&&
(\Omega_\Lambda+6\eta-1)(3\eta(4\eta-1)+(3\eta-1)\Omega_\eta)\Big),
\end{eqnarray}
plotted in Fig. \ref{figv1} versus redshift parameter. We see that
the square of sound speed is always negative during the evolution of
Universe meaning that the non-interacting HDE in Rasttal theory is
classically unstable, a behavior obtained in the framework of
standard cosmology in which HDE plays the role of dark energy
\cite{stab}.

\begin{figure}[!]
\begin{center}
    \includegraphics[width=8cm]{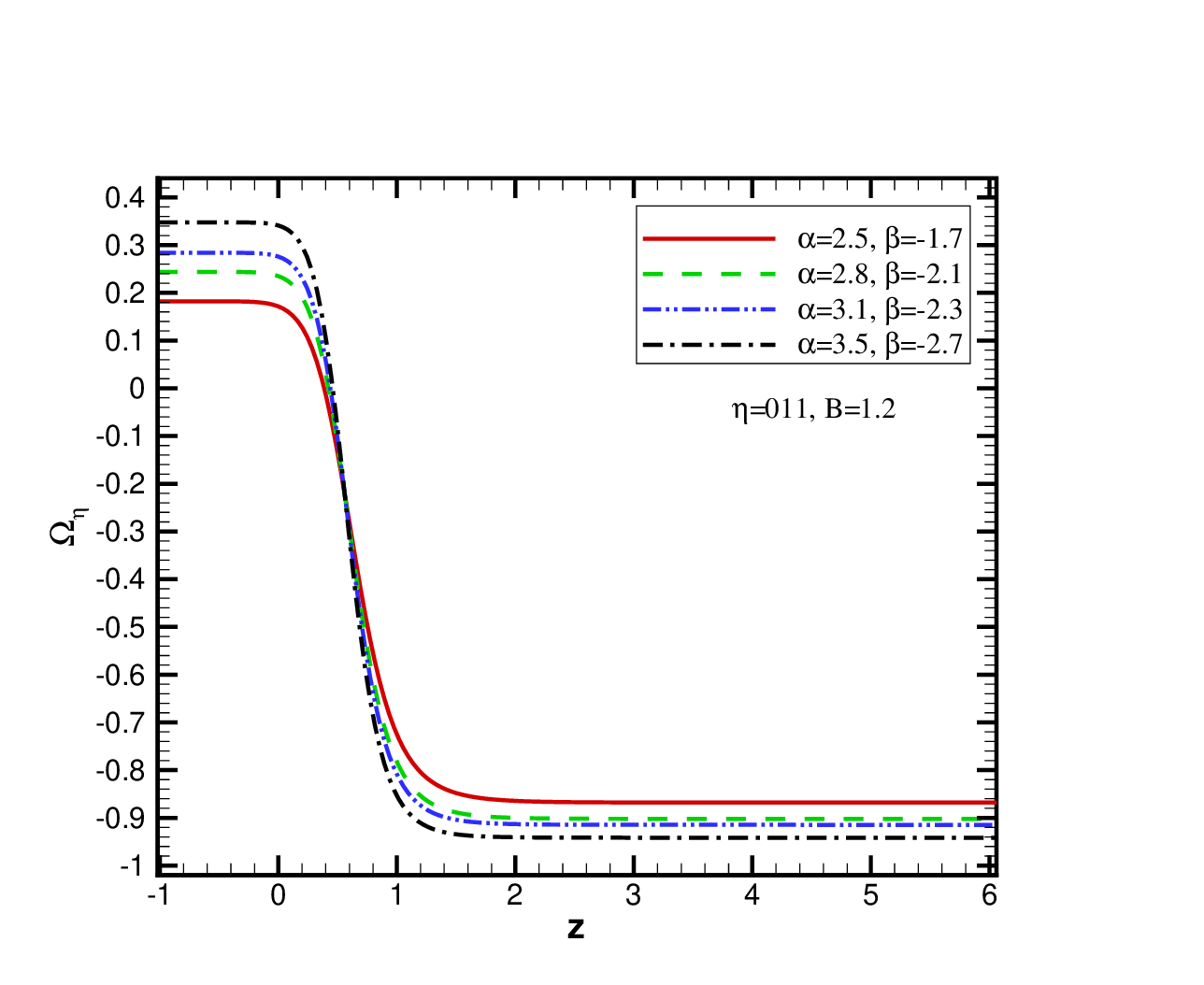}
    \caption{ $\Omega_\eta$ versus $z$ for interacting HDE in Rastall theory.
    we have taken $ \eta=0\cdot 11 $ and some values of $ B $,  $\alpha$ and $ \beta $.}\label{Omega_e2}
\end{center}
\end{figure}
\begin{figure}[htp]
\begin{center}
    \includegraphics[width=8cm]{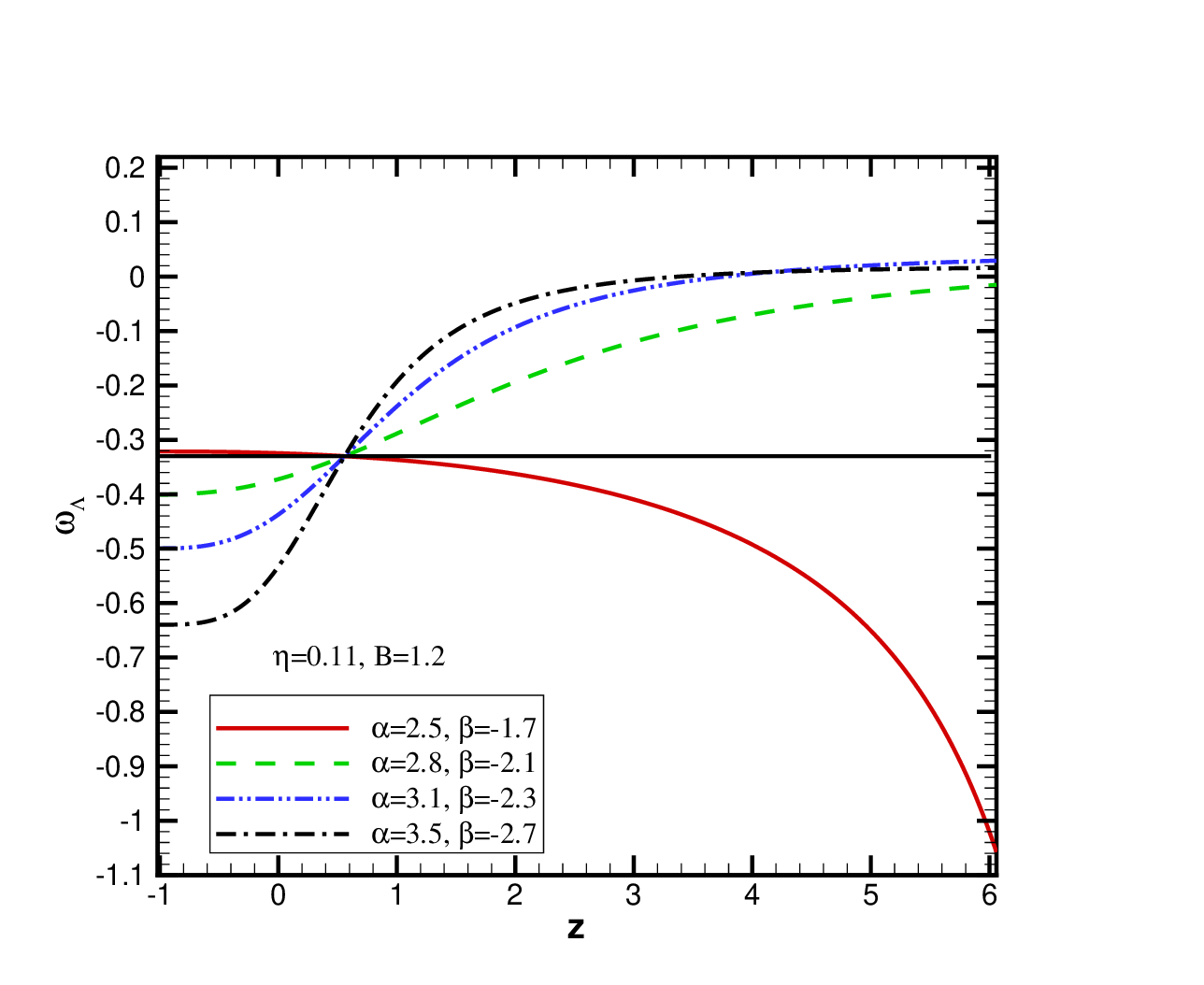}
    \includegraphics[width=7.5cm]{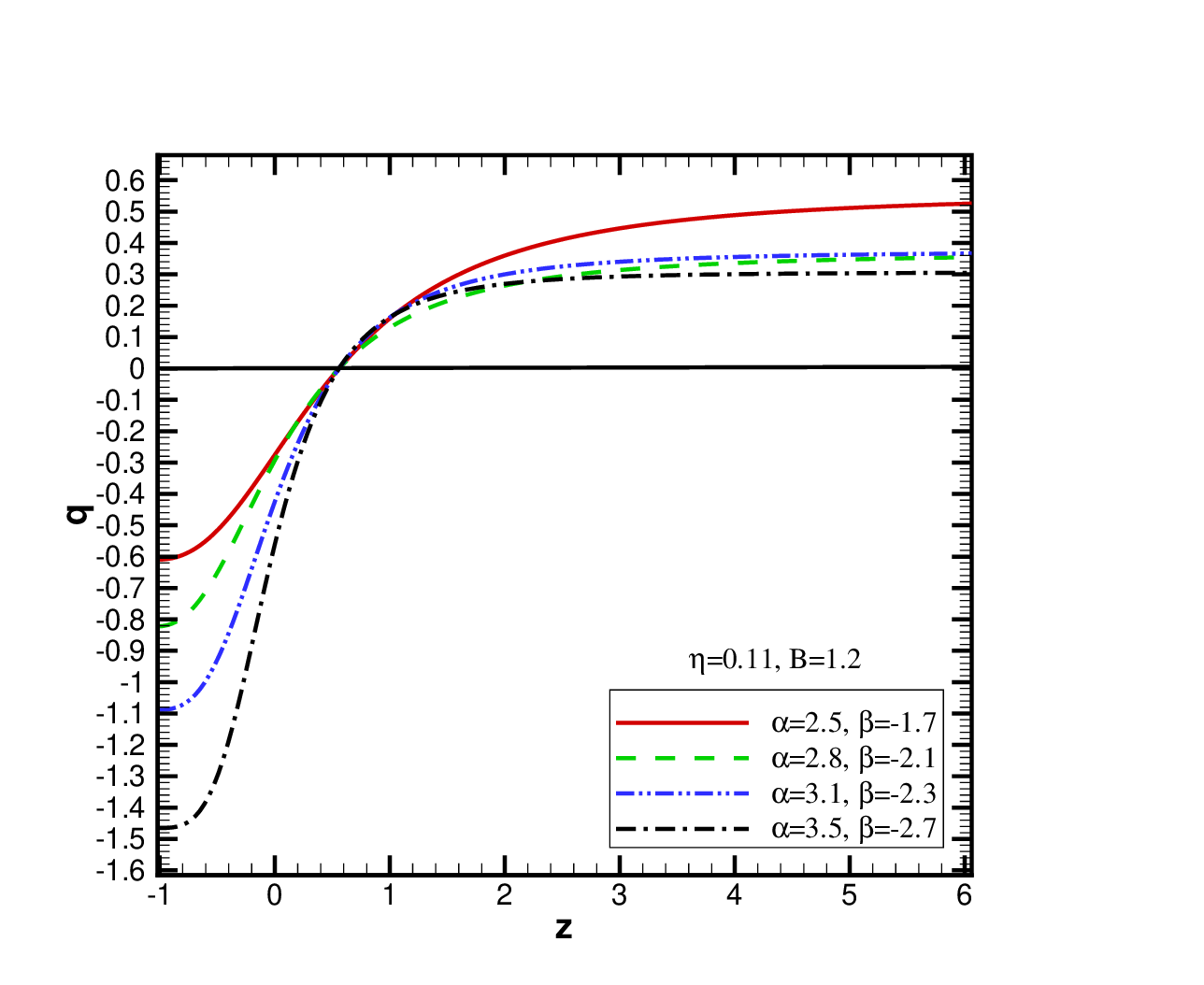}
    \caption{The evolution of $ \omega_\Lambda $ and $ q $
    with respect to$ z $ for interacting HDE in Rastall theory.}\label{figw2}
\end{center}
\end{figure}


\subsection*{An interacting Case}

In the presence of mutual interaction between DM and HDE , the
conservation equations are given as \cite{Q1,Q2,int,obser}

\begin{eqnarray}\label{Qcont2}
&&(\frac{3\eta-1}{4\eta-1})\dot{\rho}_m+3H\rho_m=Q,\\\nonumber
&&(\frac{3\eta-1}{4\eta-1})\dot{\rho}_\Lambda+(\frac{3\eta}{4\eta-1})\dot{p}_\Lambda+3H(\rho_\Lambda+p_\Lambda)=-Q,
\end{eqnarray}

\noindent where $ Q $ denotes the interaction term and we assume
that it has the form $Q=H(\alpha\rho_m+\beta\rho_D)$, in which $
\alpha $ and $ \beta $ are unknown coupling constants \cite{int}.
Using Eqs.~(\ref{hder}),~(\ref{Qcont2}), and the time derivative
of Eq. (\ref{friedman3}), we reach

\begin{eqnarray}
\frac{\ddot{H}}{H^3}&=&\frac{4\eta-1}{2\eta(1-6\eta)}
\Big[\frac{\Omega_\eta(\alpha-3)(6\eta-1)}{1-3\eta}\nonumber\\
&+&(\Omega_\Lambda+6\eta-1)\Big(\frac{\Omega_\eta}{\eta}+\frac{(\alpha-3)(4\eta-1)}{1-3\eta}\Big)\nonumber
\\&+&\frac{(4\eta-1)\beta\Omega_\Lambda}{3\eta-1}\Big],
\end{eqnarray}
used in order to write
\begin{eqnarray}\label{dOmega2}
\Omega_\eta^\prime&=&\frac{4\eta-1}{(1-6\eta)}
\Big[\frac{\Omega_\eta(\alpha-3)(6\eta-1)}{1-3\eta}\nonumber\\
&+&(\Omega_\Lambda+6\eta-1)\Big(\frac{\Omega_\eta}{\eta}+\frac{(\alpha-3)(4\eta-1)}{1-3\eta}\Big)\nonumber
\\&+&\frac{(4\eta-1)\beta\Omega_\Lambda}{3\eta-1}\Big]-\frac{\Omega_\eta^2}{\eta}.
\end{eqnarray}

\noindent The evolution of $ \Omega_\eta $ against redshift $ z $
has been plotted in Fig. \ref{Omega_e2} for the initial condition
$ \Omega_\Lambda(z=0)=0\cdot73 $. We have also plotted the
evolution of the EoS and deceleration parameters against redshift
$ z $ for some values of parameters in Fig.~\ref{figw2}.
From this figure, one can clearly see the transition redshift $ z_t
$ lies within the interval $ 0.5<z<0.8 $.

\begin{figure}[!]
\begin{center}
    \includegraphics[width=8cm]{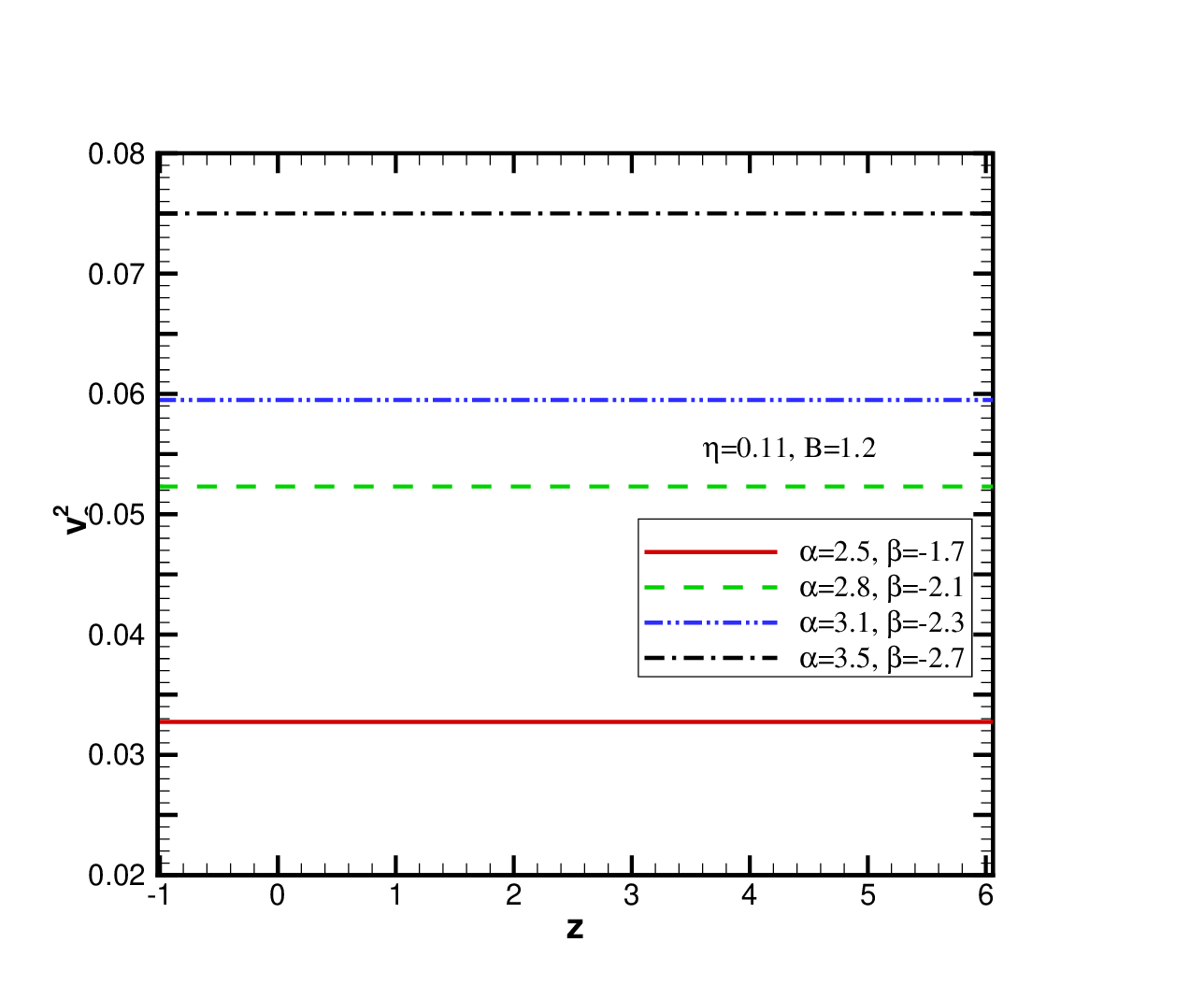}
    \caption{ $v_s^2$ versus $z$ for interacting HDE in Rastall theory
    for $ \eta=0\cdot 11 $ and some values of $ B $, $\alpha$ and $ \beta $.}\label{figv2}
\end{center}
\end{figure}

By taking the time derivative of Eq.~(\ref{w1}) and combining the
result with Eqs.~(\ref{hder}) and (\ref{dOmega2}), we reach at
\begin{eqnarray}\label{v_s2}
v_s^2&=&\frac{4\eta-1}{B}+\frac{4\eta-1}{3B(1-6\eta)\Omega_\eta}\Big[(3-\alpha)(6\eta-1)\Omega_\eta\nonumber\\
&+&(\Omega_\Lambda+6\eta-1)\Big((3-\alpha)(4\eta-1)+\frac{(3\eta-1)\Omega_\eta}{\eta}\Big)\nonumber\\&+&
(4\eta-1)\beta\Omega_\Lambda\Big].
\end{eqnarray}
The evolution of $ v_s^2 $ versus redshift parameter is plotted in
Fig. \ref{figv2} showing that the interacting HDE model in Rastall
theory, unlike the non-interacting case, is classically stable.

\subsection{A new approach}

Here, we assume DE is a combination of the Rastall term and vacuum
energy. In this case, Eqs.~(\ref{friedman3}) can be rewritten as

\begin{eqnarray}\label{friedman5}
&&3H^2=\kappa(\rho_m+\rho_D),\nonumber\\
&&H^2+\frac{2}{3}\dot{H}=-\frac{\kappa}{3}p_D,
\end{eqnarray}

\noindent where

\begin{eqnarray}\label{rhoeff}
&&\rho_D=\rho_\Lambda+\frac{6\eta(6\eta-1)}{\kappa_G(4\eta-1)}(2H^2+\dot{H}),\nonumber\\
&&p_D=p_\Lambda-\frac{6\eta(6\eta-1)}{\kappa_G(4\eta-1)}(2H^2+\dot{H}).
\end{eqnarray}

\noindent It is easy to see that the standard cosmology is restored
at the appropriate limit $\eta\to0$. Indeed, we wrote the Friedmann
equations in a form easily comparable with the standard Friedmann
equations. In this case, by defining critical density as
$\rho_c=\frac{3H^2(6\eta-1)}{\kappa_G(4\eta-1)}$, and using
Eq.~(\ref{friedman5}), one can find

\begin{equation}\label{friedman6}
\Omega_m+\Omega_D=1,
\end{equation}

\noindent where

\begin{eqnarray}\label{OmegaD}
&&\Omega_m=\frac{\rho_m}{\rho_c}=\frac{\kappa\rho_m}{3H^2},~~~\Omega_\Lambda=\frac{\rho_\Lambda}{\rho_c}=B\nonumber\\
&&\Omega_D=\Omega_\Lambda+4\eta+2\eta\frac{\dot{H}}{H^2}.
\end{eqnarray}

\noindent The use of Eq.~(\ref{friedman5}) leads also to

\begin{equation}\label{wD}
\omega_D=\frac{-1}{\Omega_D}\Big(1+\frac{2\dot{H}}{3H^2}\Big).
\end{equation}
Taking the time derivative of Eq.~(\ref{OmegaD}), we obtain

\begin{equation}\label{dOmegaD}
\dot{\Omega}_D=2\eta\Big(\frac{\ddot{H}}{H^2}-2\frac{\dot{H}^2}{H^3}\Big).
\end{equation}

\noindent For the limiting case $ \eta\rightarrow 0 $, the
equation of motion of HDE in standard cosmology $
(\Omega_D=\Omega_\Lambda=const) $, as a desired result, is
restored.

Substituting Eq.~(\ref{OmegaD}) into (\ref{wD}), one can obtain
the EoS as
\begin{equation}\label{w3}
\omega_D=\frac{-1}{3\eta\Omega_D}\Big(\Omega_D-\Omega_\Lambda-\eta\Big).
\end{equation}
Using Eq.~(\ref{OmegaD}), one also finds
\begin{equation}\label{q3}
q=-1-\frac{\dot{H}}{H^2}=\frac{-1}{2\eta}(\Omega_D-\Omega_\Lambda-2\eta).
\end{equation}
\noindent Hence, it is obvious that at $ \eta\rightarrow 0 $ $
(\Omega_D=\Omega_\Lambda) $ limit, the EoS and deceleration
parameters for the HDE in standard cosmology are recovered. It is
also useful to mention that Eq.~(\ref{w3}) and~(\ref{q3}) are the
same for both non-interacting and interacting cases.
\begin{figure}[!]
\begin{center}
    \includegraphics[width=8cm]{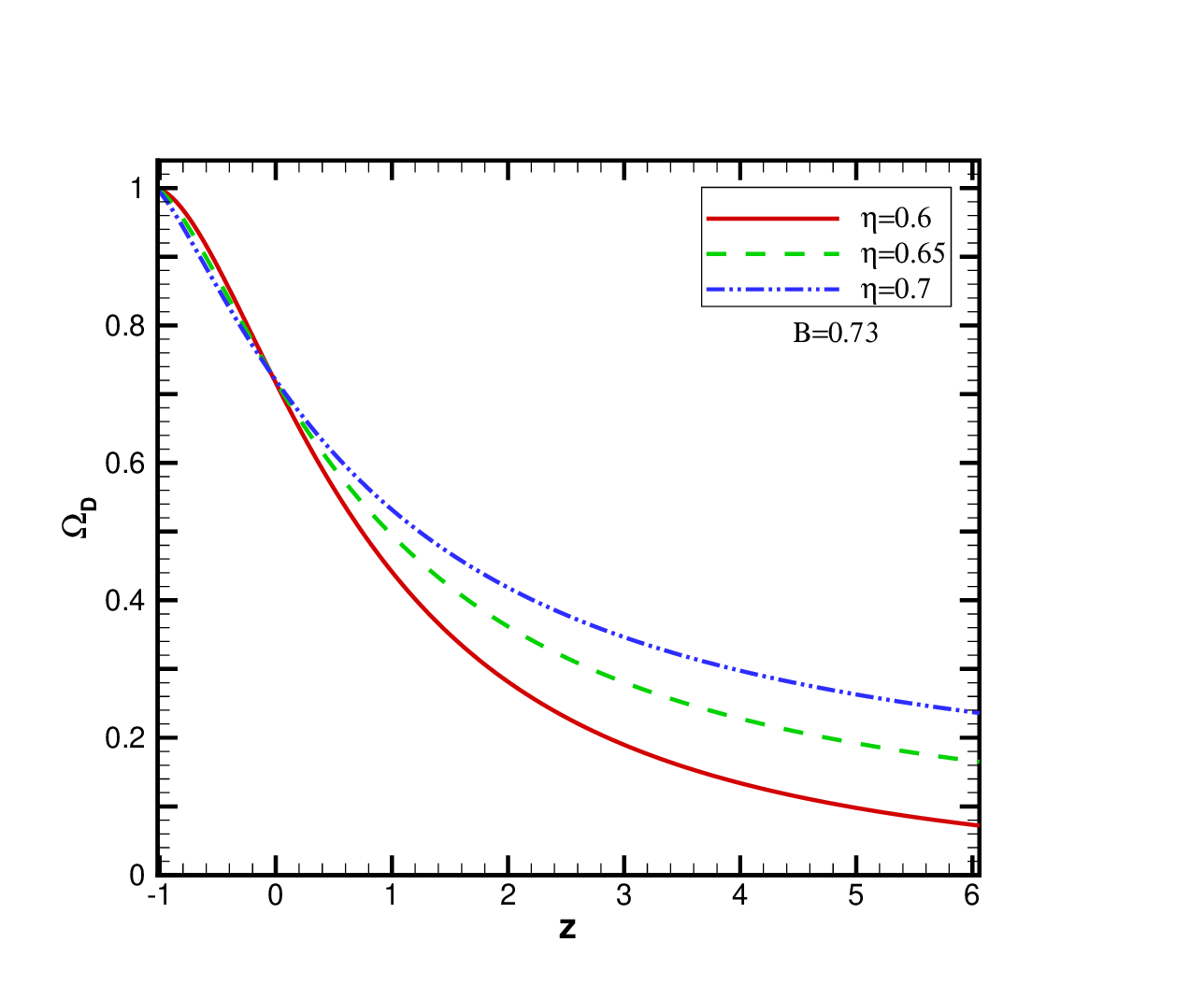}
    \caption{ $\Omega_D$ versus $z$ for non-interacting HDE model for $\Omega_{D}(z=0)=0\cdot73$,
    $ B=0\cdot73 $ and some values of $\eta$.}\label{Omega_D1}
\end{center}
\end{figure}

\begin{figure}[htp]
\begin{center}
    \includegraphics[width=8cm]{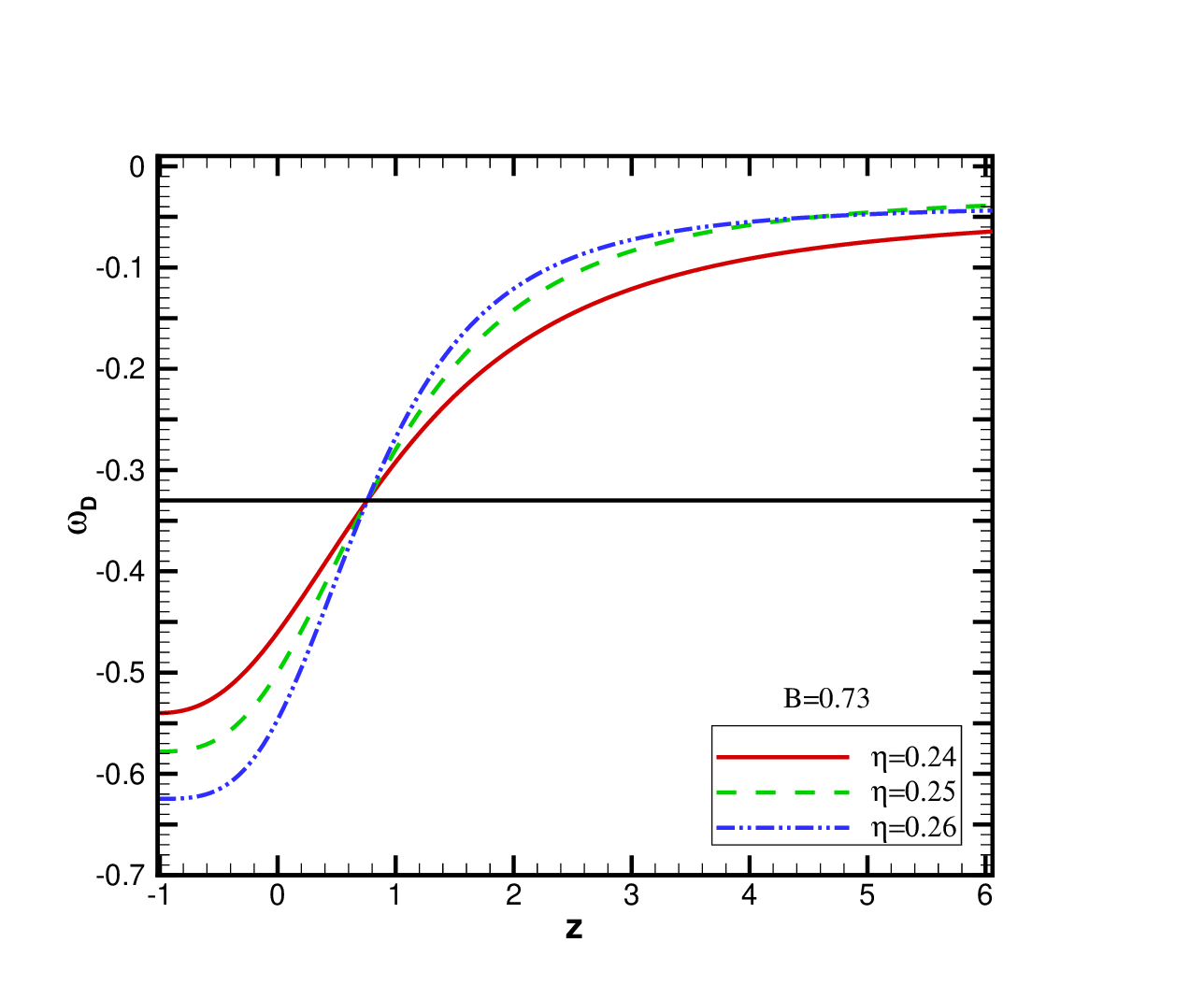}
    \includegraphics[width=7.5cm]{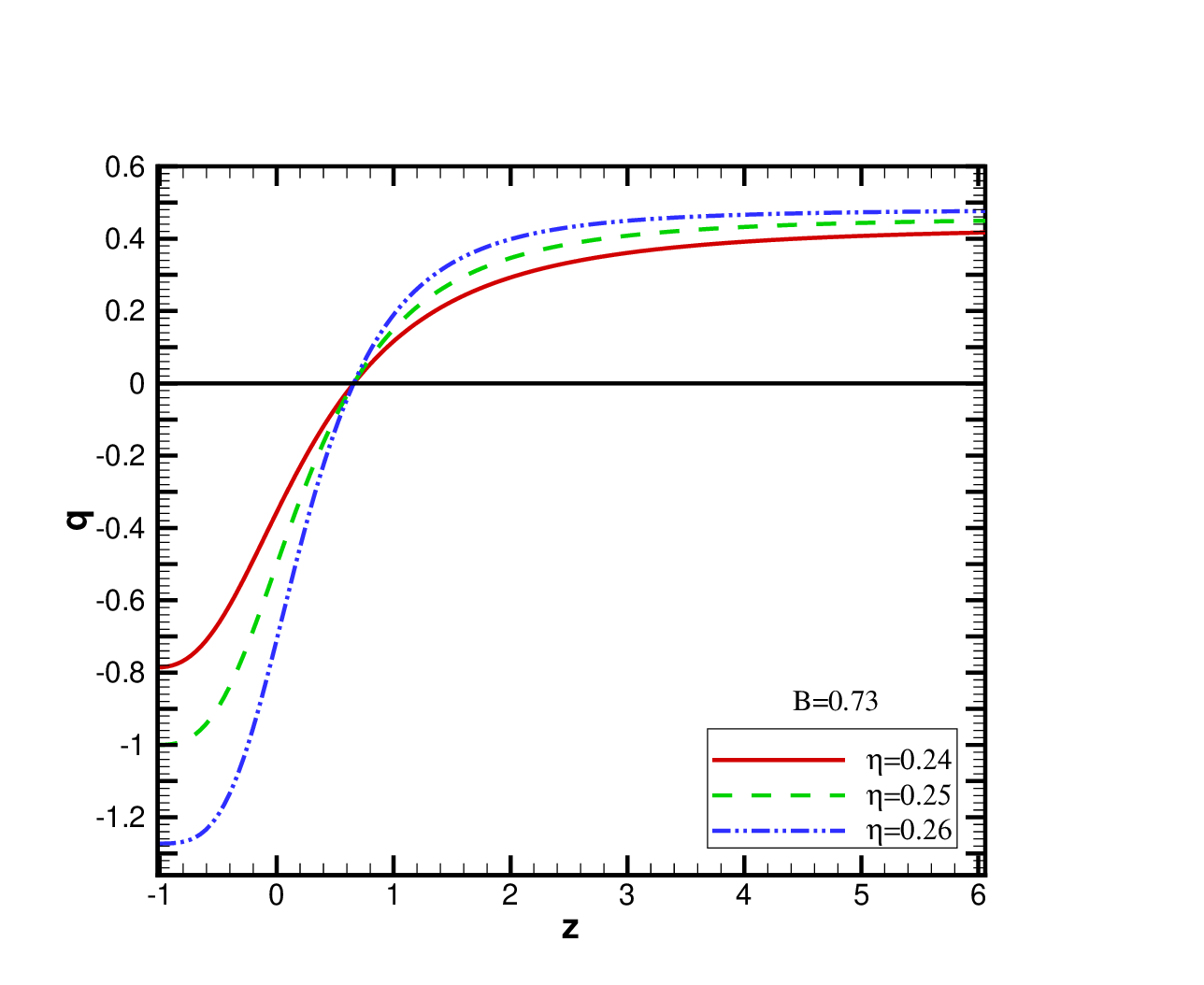}
    \caption{The evolution of $ \omega_D $ and $ q $ versus $ z $
    for the non-interacting HDE in Rastall theory for $ B=0\cdot73 $ and some values of $ \eta $.}\label{figw3}
\end{center}
\end{figure}
\begin{figure}[!]
\begin{center}
    \includegraphics[width=8cm]{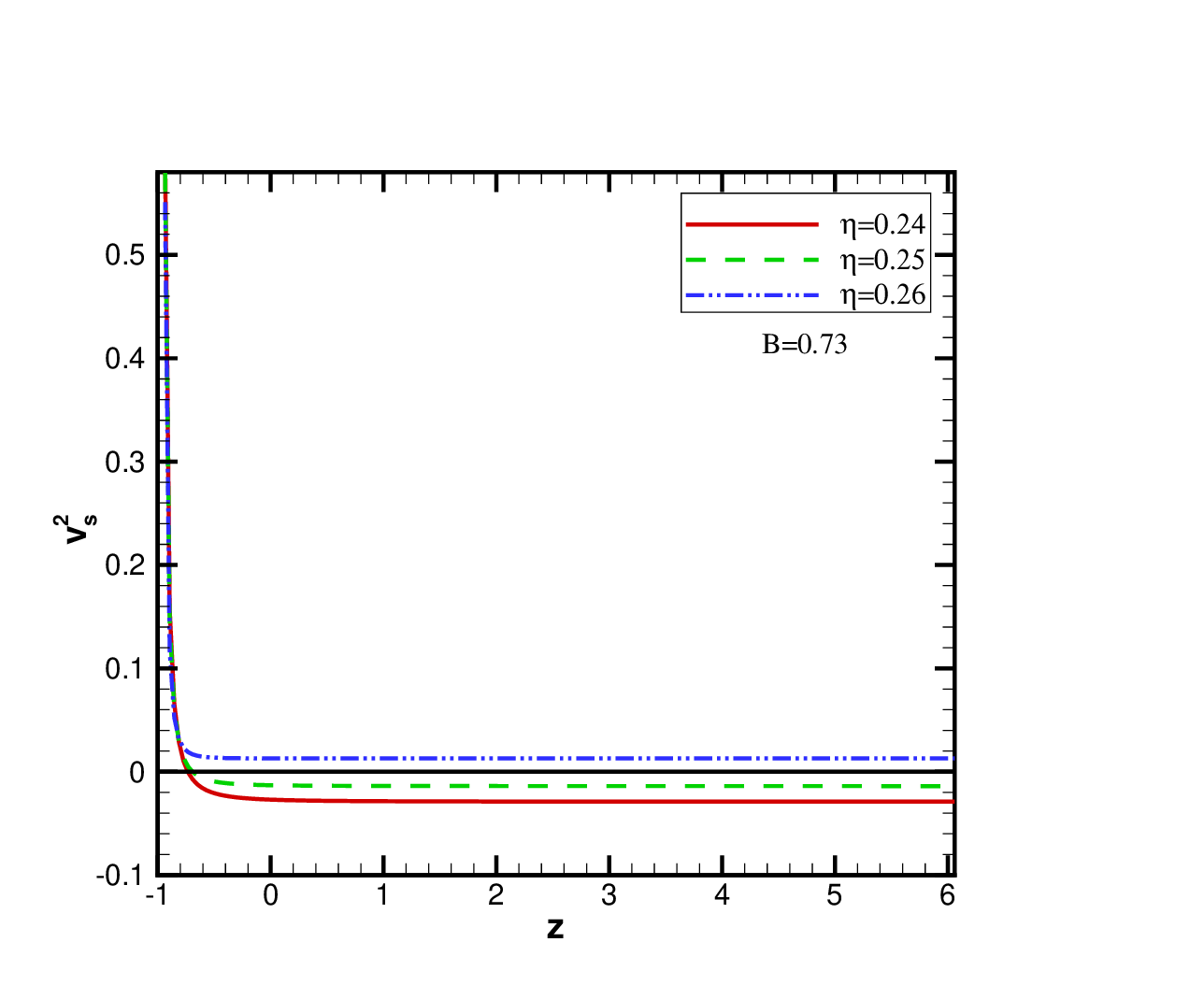}
    \caption{ $v_s^2(z)$ for non-interacting HDE for
    some values of $\alpha$ and  $\beta$ .}\label{figv3}
\end{center}
\end{figure}

\subsection*{Non-interacting Case}
\noindent Combining
Eqs.~(\ref{cont1}),~(\ref{hder}),~(\ref{friedman6})
and~(\ref{OmegaD}) with the time derivative of the first Friedmann
equation~(\ref{friedman5}), we arrive

\begin{eqnarray}\label{ddH2}
\frac{\ddot{H}}{H^3}&=&\frac{3(4\eta-1)(\Omega_D-1)}{2\eta(1-3\eta)}\nonumber\\
&-&\frac{1}{2\eta^2}(\Omega_\Lambda+4\eta-1)(\Omega_D-\Omega_\Lambda-4\eta).
\end{eqnarray}

\noindent Inserting Eqs.~(\ref{OmegaD}) and (\ref{ddH2}) into
(\ref{dOmegaD}), we can also obtain the evolution of dimensionless
DE density parameter as

\begin{equation}\label{dOmegaD2}
\Omega_D^{\prime}=(\Omega_D-1)\Big(\frac{3(4\eta-1)}{1-3\eta}-\frac{\Omega_D-\Omega_\Lambda-4\eta}{\eta}\Big).
\end{equation}

\noindent The evolution of $ \Omega_D $ against redshift $ z $ has
been plotted in Fig.\ref{Omega_D1} for the initial condition $
\Omega_D(z=0)=0.73 $ and  some values of $ \eta $, addressing us
that $ \Omega_D\rightarrow 0 $  and $ \Omega_D\rightarrow 1 $ at
the early time and late time, respectively. Moreover, the behavior
of $ \omega_D(z) $ and $ q(z) $ are shown numerically in Fig.
\ref{figw3} for some values of the parameter $ \eta $ which
implies that Universe has experienced a transition at the redshift
$ z\approx 0\cdot6 $, and the EoS parameter can not cross the
phantom divide $ (\omega_D=-1) $. Combining
Eqs.~(\ref{hder}),~(\ref{dOmegaD3}),~(\ref{ddH4}), and the time
derivative of Eq.~(\ref{w3}) with Eq.~(\ref{v_s}), we get
\begin{eqnarray}
&&v_s^2=\frac{dp_D}{d\rho_D}=\frac{-1}{3\eta\Omega_D}\Big(\Omega_D-\Omega_\Lambda-\eta\Big)-\\&&\nonumber
\frac{(\Omega_\Lambda+\eta)(1-\Omega_D)}{3\Omega_D}\Big(\frac{3(4\eta-1)}{1-3\eta}-\frac{\Omega_D-\Omega_\Lambda-4\eta}{\eta}\Big)\nonumber\\&&
\times\Big(\frac{1-3\eta}{3\eta(4\eta-1)(\Omega_D-1)+(1-3\eta)(\Omega_D-\Omega_\Lambda-4\eta)}\Big),\nonumber
\end{eqnarray}
plotted in Fig. \ref{figv3} for non-interacting case. It is
therefore seen that the stability of this case depends on the value
of Rastall parameter so that there exists a critical value for this
parameter that lies within the interval $0.25<\eta_c<0.26$ and
separates the stable and unstable regimes. Therefore we have
classical stability (un-stability) for $\eta>\eta_c$
($\eta<\eta_c$).


\subsection*{An interacting Case}

\begin{figure}[!]
\begin{center}
    \includegraphics[width=8cm]{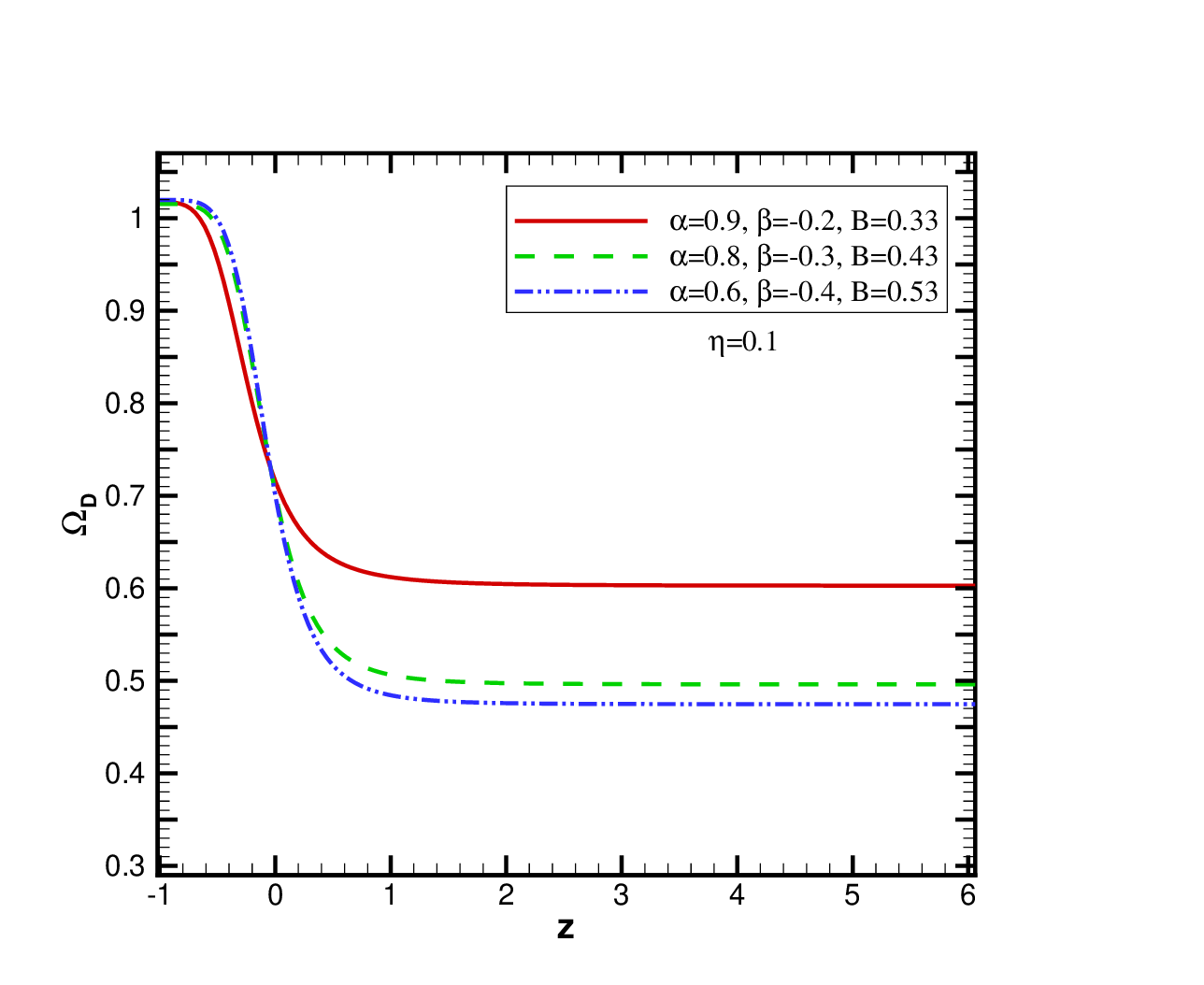}
    \caption{ $\Omega_D$ versus $z$ for interacting HDE for some values of $ B $, $\alpha$
    and  $\beta$.}\label{figOmega2}
\end{center}
\end{figure}

Taking the time derivative of Eq.(\ref{friedman5}) along with using
Eqs.~(\ref{hder}),~(\ref{Qcont2}),~(\ref{friedman6})
and~(\ref{OmegaD}), we reach at
\begin{eqnarray}\label{ddH4}
\frac{\ddot{H}}{H^3}&=&\frac{4\eta-1}{2\eta(1-3\eta)}((\alpha-3)(1-\Omega_D)+\beta\Omega_\Lambda)
\nonumber\\&-&\frac{1}{2\eta^2}(\Omega_\Lambda+4\eta-1)(\Omega_D-\Omega_\Lambda-4\eta).
\end{eqnarray}
Inserting Eq.~(\ref{ddH4}) into (\ref{dOmegaD}), one gets
\begin{eqnarray}\label{dOmegaD3}
\Omega_D^{\prime}&=&\frac{4\eta-1}{1-3\eta}((\alpha-3)(1-\Omega_D)+\beta\Omega_\Lambda)
\nonumber\\&-&\frac{1}{\eta}(\Omega_D-1)(\Omega_D-\Omega_\Lambda-4\eta),
\end{eqnarray}
plotted as a function of $ z $ in Fig. \ref{figOmega2} for the
initial condition $ \Omega_D(z=0)=0.73 $. In this manner, while we
have $ \Omega_D\rightarrow 1 $ at the late time ($ z\rightarrow-1$),
$ 0.2<\Omega_D<0.3 $ at the early time ($ z\rightarrow \infty$), in
agreement with Eq.~(\ref{OmegaD}) where the DE density parameter is
a combination of the Rastall term and vacuum energy.

\begin{figure}[htp]
\begin{center}
    \includegraphics[width=8cm]{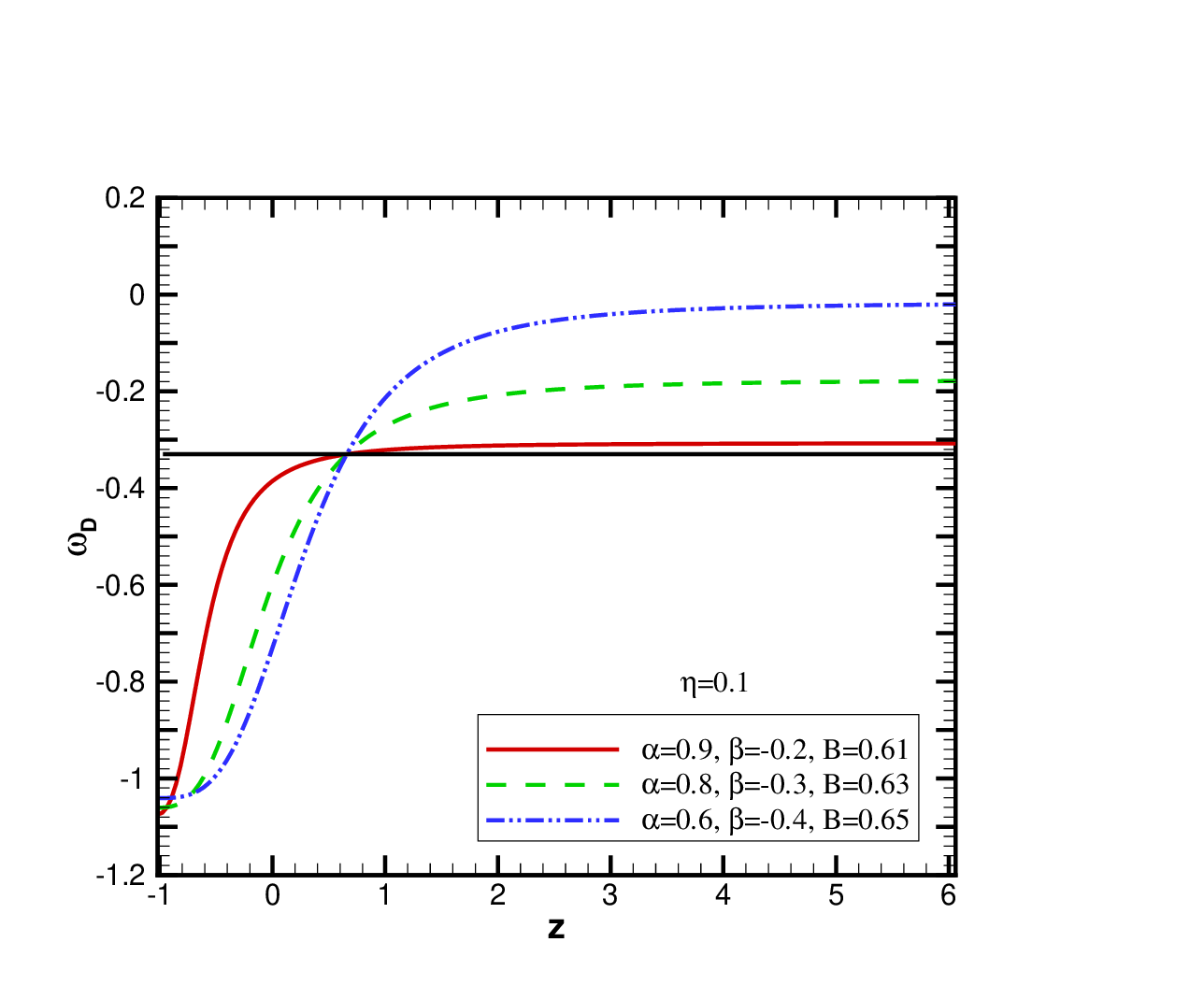}
    \includegraphics[width=7.5cm]{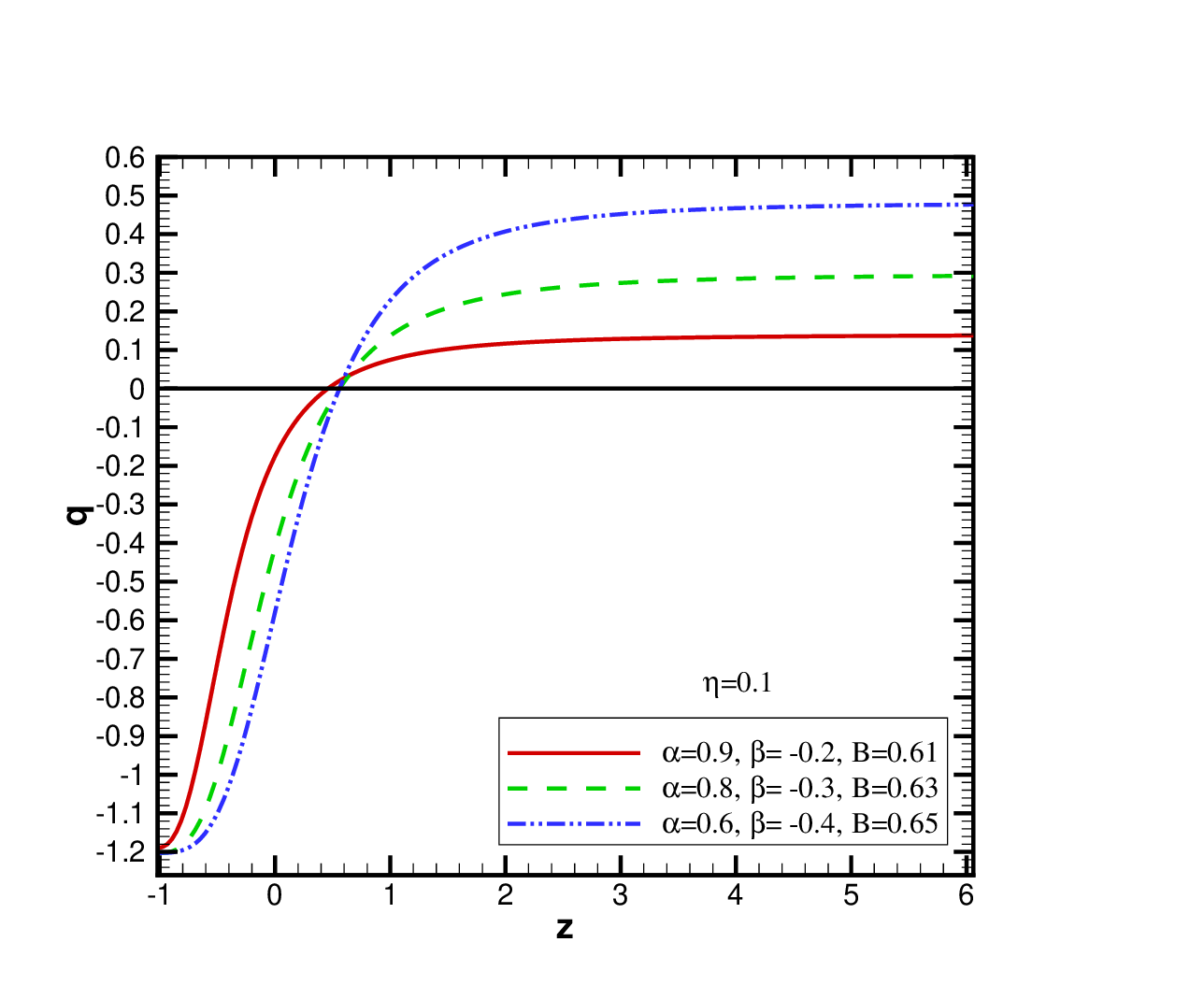}
    \caption{The evolution of the $ \omega_D $ and $ q $ parameters
    with respect to the $ z $ for interacting HDE in Rastall theory.}\label{figw4}
\end{center}
\end{figure}

Figure. \ref{figw4} includes the evolution of $\omega_\Lambda(z)$
and $ q(z) $ for some values of $ \alpha $ and $ \beta $ parameters,
and indicates that there is a transition from the deceleration phase
to the accelerated phase in the interval $ 0.5<z_t<0.9$. It is also
seen that, in this case, the interacting HDE model behaves like the
phantom DE at the late time.

\begin{figure}[!]
\begin{center}
    \includegraphics[width=8cm]{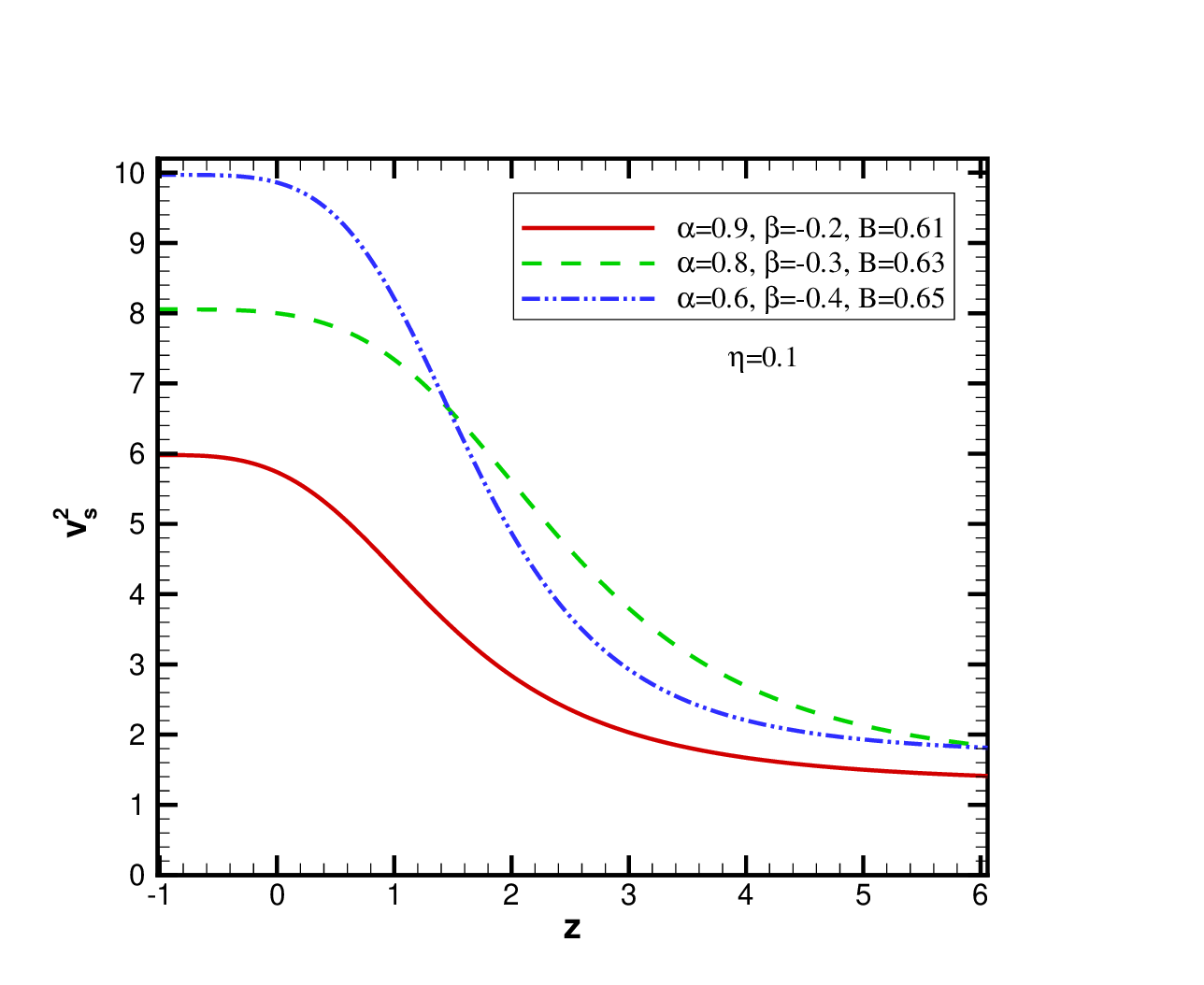}
    \caption{ $v_s^2(z)$ for the interacting HDE for
    some values of $ B $, $\alpha$ and  $\beta$ .}\label{figv4}
\end{center}
\end{figure}


\begin{figure}[htp]
\begin{center}
    \includegraphics[width=8cm]{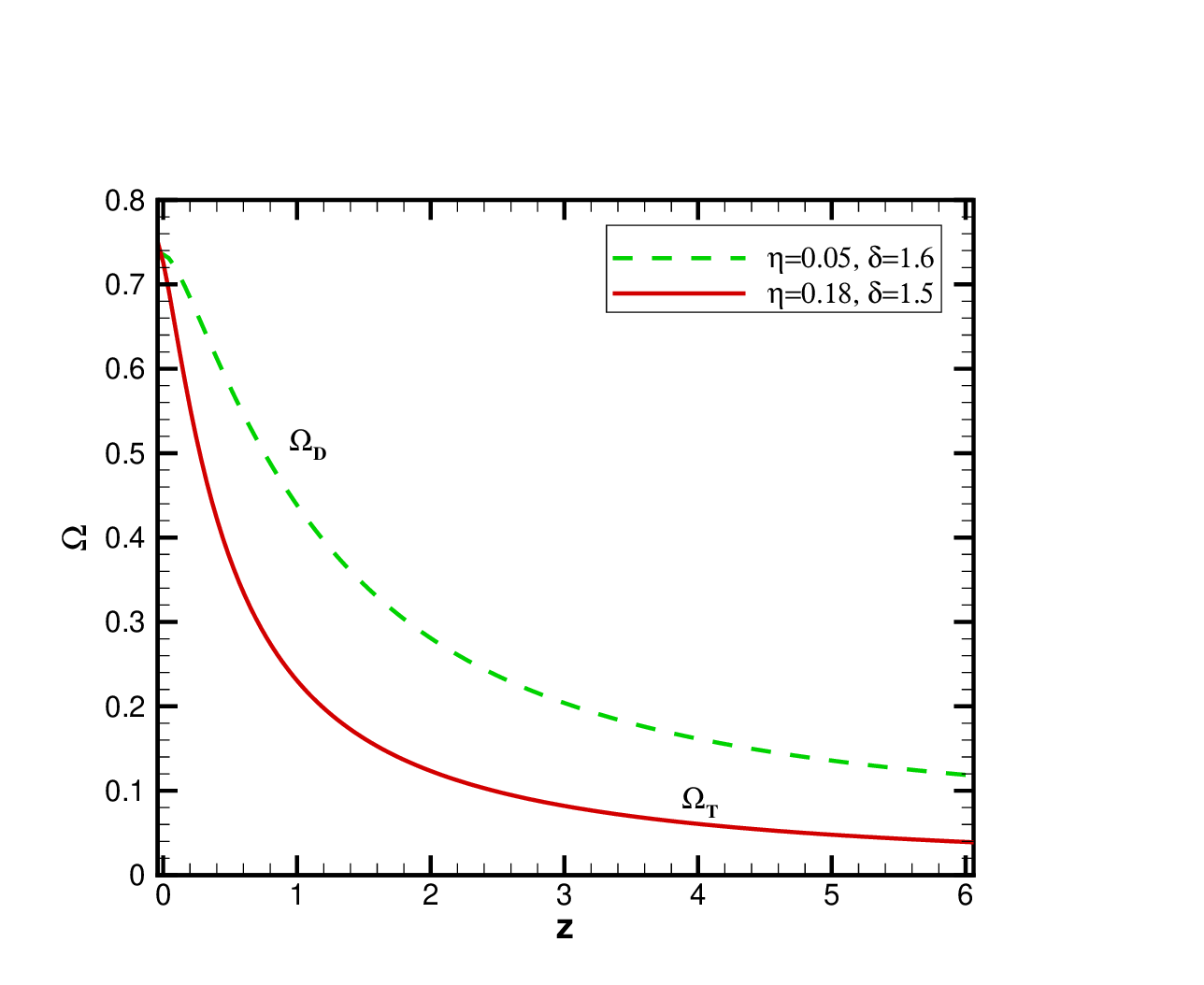}
    \caption{The $ \Omega_T $ and $ \Omega_D $ against $ z $
    for the non-interacting THDE model for the common (solid line) and new (dash line)
    approach in Rastall theory.}\label{figOmegaT1}
\end{center}
\end{figure}
\begin{figure}[htp]
\begin{center}
    \includegraphics[width=8cm]{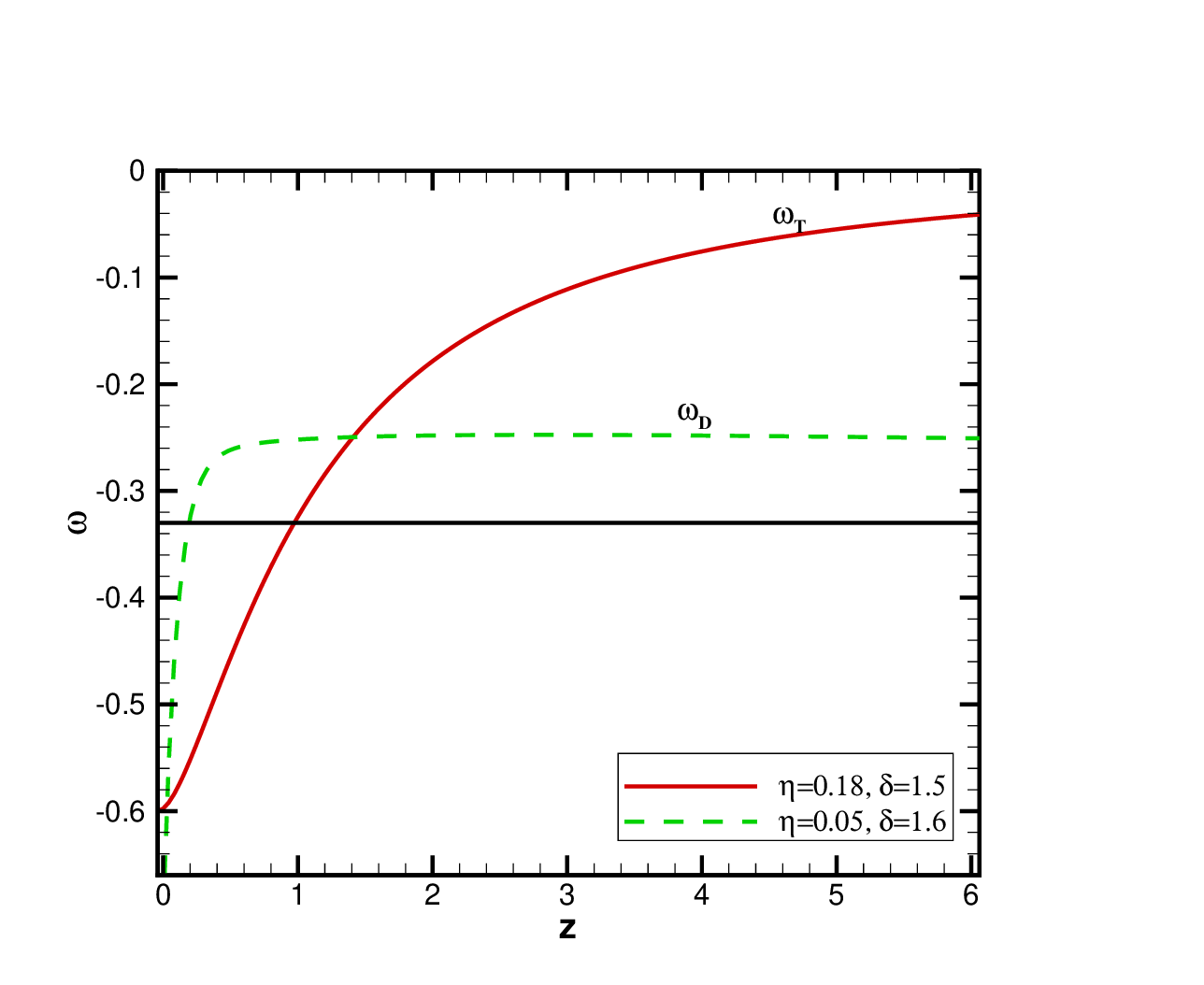}
    \includegraphics[width=7.5cm]{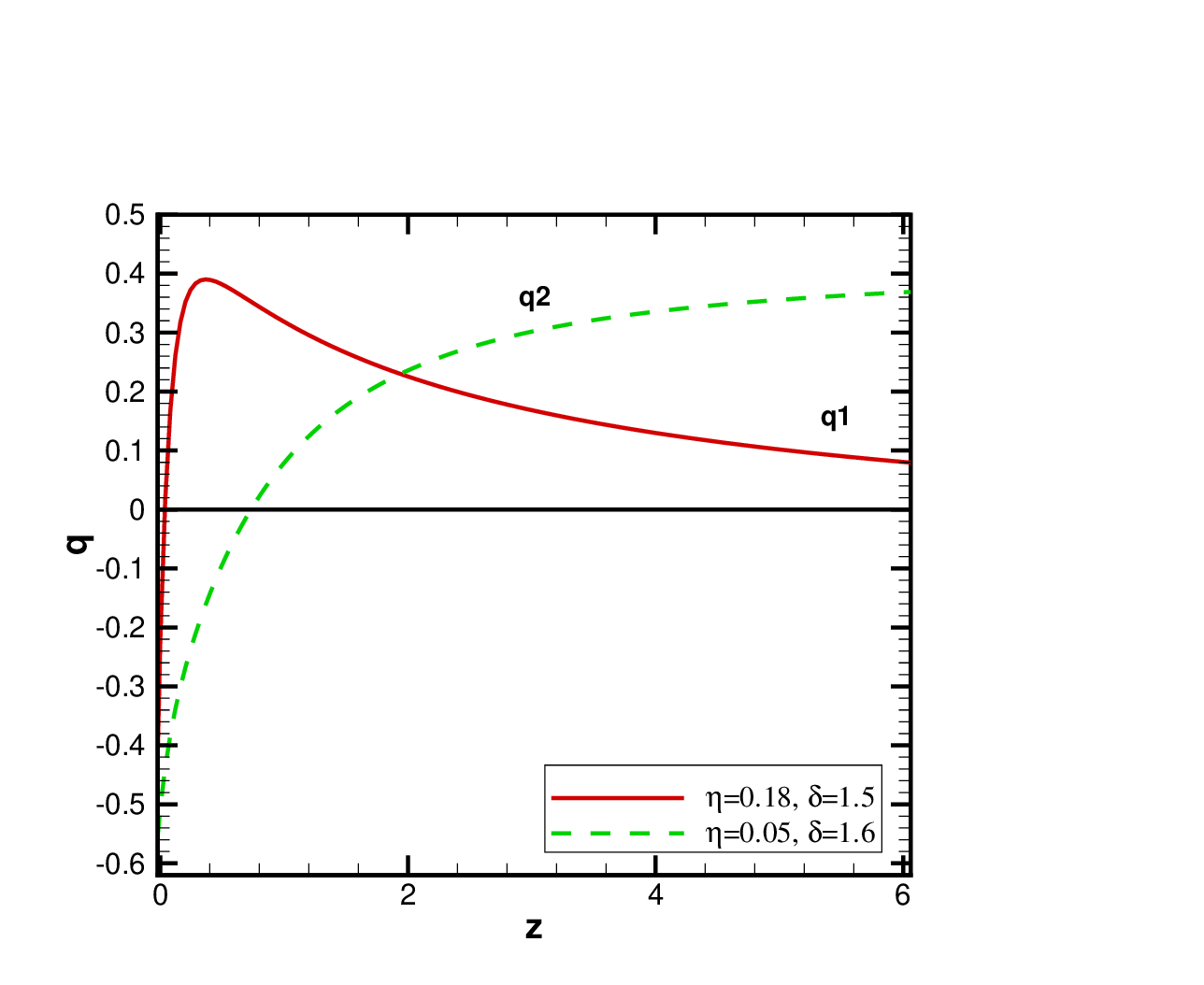}
    \caption{The  $ \omega $ and  $ q $ against $ z $
    for the non-interacting THDE medel for the common (solid line) and new (dash line)
    approach in Rastall theory.}\label{figwT1}
\end{center}
\end{figure}


\section{THDE in Rastall gravity}

Following \cite{Tavayef}, THDE density in Rastall theory is
obtained as

\begin{eqnarray}\label{THDE}
\rho_T=\frac{3B}{8\pi G}(\frac{6\eta-1}{4\eta-1})H^{4-2\delta},
\end{eqnarray}

\noindent where $\delta$ and $B$ are unknown parameters.
\subsection{The common approach}

\noindent Using definition~(\ref{omega}), we can write THDE
dimensionless density parameter and its evolution as

\begin{eqnarray}\label{OmegaT1}
\Omega_T=B\Big(\frac{6\eta-1}{4\eta-1}\Big)H^{2-2\delta},
\end{eqnarray}
and
\begin{eqnarray}\label{dOmegaT1}
\Omega_T^{\prime}=\frac{(1-\delta)\Omega_T\Omega_\eta}{\eta},
\end{eqnarray}
respectively. One can also use
Eqs.~(\ref{friedman3}),~(\ref{omega}) and~(\ref{THDE}) in order to
obtain EoS and deceleration parameters as
\begin{equation}\label{wT1}
\omega_T=\frac{(4\eta-1)\Omega_T}{B^2(6\eta-1)}\Big(\frac{(3\eta-1)\Omega_\eta}{\eta}+4\eta-1\Big),
\end{equation}
and
\begin{equation}\label{qT1}
q=-1-\frac{\Omega_\eta}{2\eta},
\end{equation}
respectively, which are, indeed, the same for both
non-interacting and interacting cases.
\subsection*{Non-interacting case}

Taking the time derivative of the first Friedmann
equation.~(\ref{friedman3}), and putting
Eqs.~(\ref{cont1}),~(\ref{omega}),~(\ref{THDE})
and~(\ref{OmegaT1}) in the result, we reach at
\begin{eqnarray}\label{dOmegaeT1}
\Omega_\eta^\prime&=&\frac{3(4\eta-1)\Omega_\eta}{1-3\eta}-\frac{\Omega_\eta^2}{\eta}\nonumber\\
&+&\frac{4\eta-1}{\eta(1-6\eta)}
\Big[\frac{3\eta(4\eta-1)}{3\eta-1}(\Omega_T+6\eta-1)\nonumber\\&+&((2-\delta)\Omega_T+6\eta-1)\Omega_\eta\Big].
\end{eqnarray}
$ \Omega_T $, $ \omega_T $ and $ q $ versus the redshift parameter
$ z $ have been plotted in Figs. \ref{figOmegaT1} and \ref{figwT1}
for non-interacting THDE in common approach. Although $
\Omega_T(z) $, $ \omega_T(z) $ and $ q(z) $ have suitable behavior
during the early to current times and can cover the current
acceleration, our study shows that proper behavior is not
obtainable for $z<0$.

\begin{figure}[htp]
\begin{center}
    \includegraphics[width=8cm]{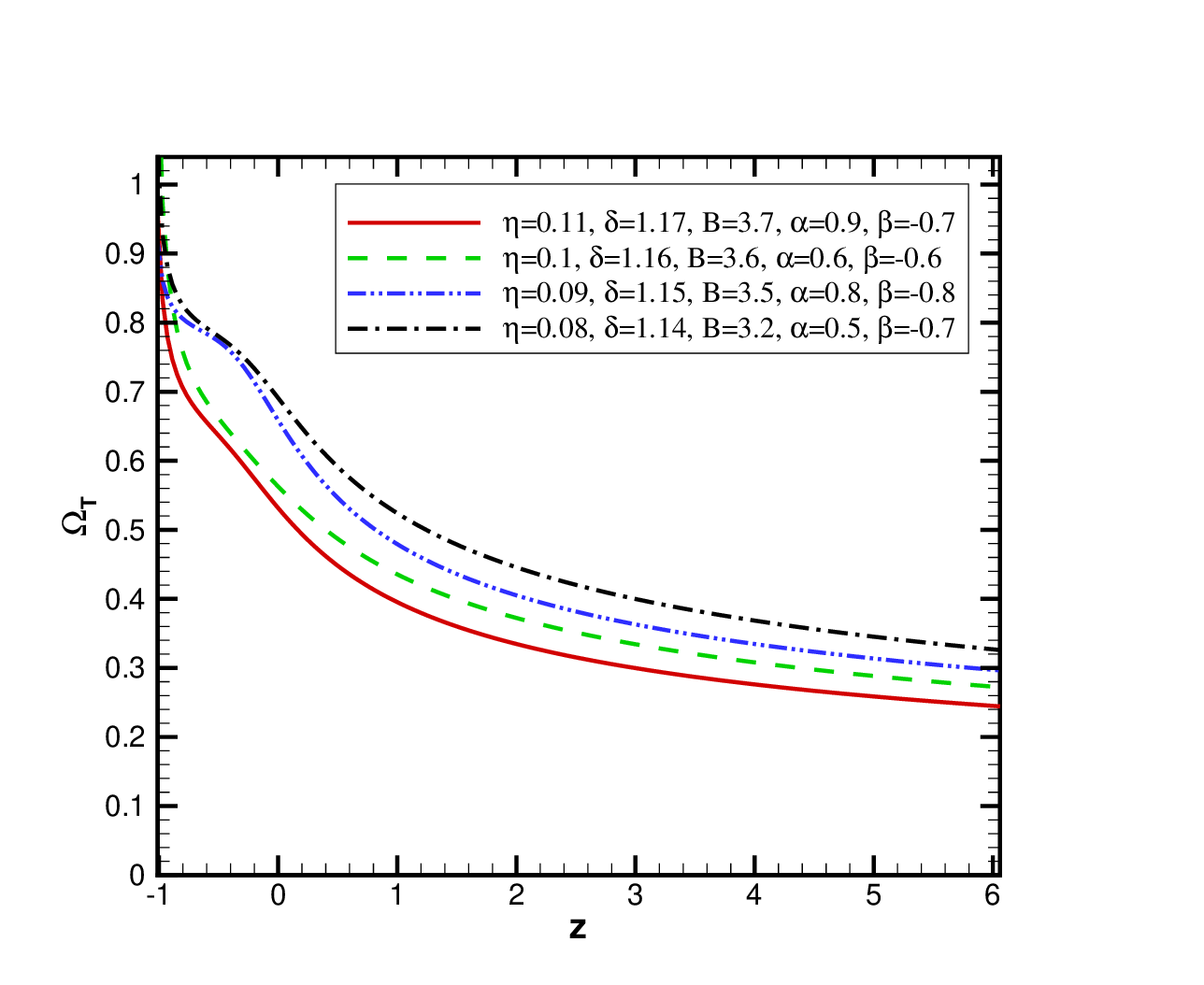}
    \caption{$ \Omega_T $ against $ z $
    for the interacting THDE model in Rastall theory. We have taken
    $ \Omega_T(z=0)=0\cdot73 $ as the initial condition.}\label{figOmegaT2}
\end{center}
\end{figure}
\begin{figure}[htp]
\begin{center}
    \includegraphics[width=8cm]{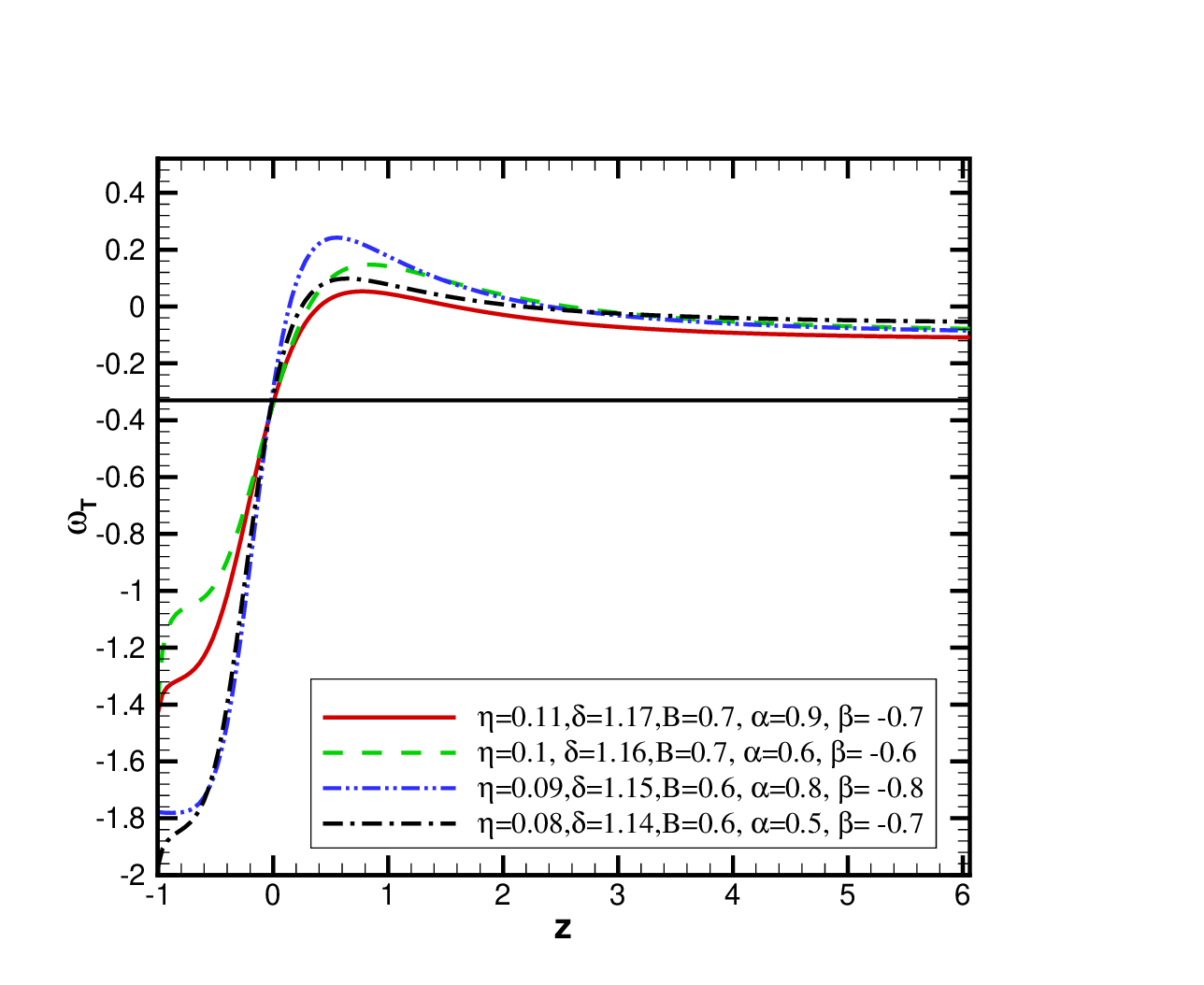}
    \includegraphics[width=7.5cm]{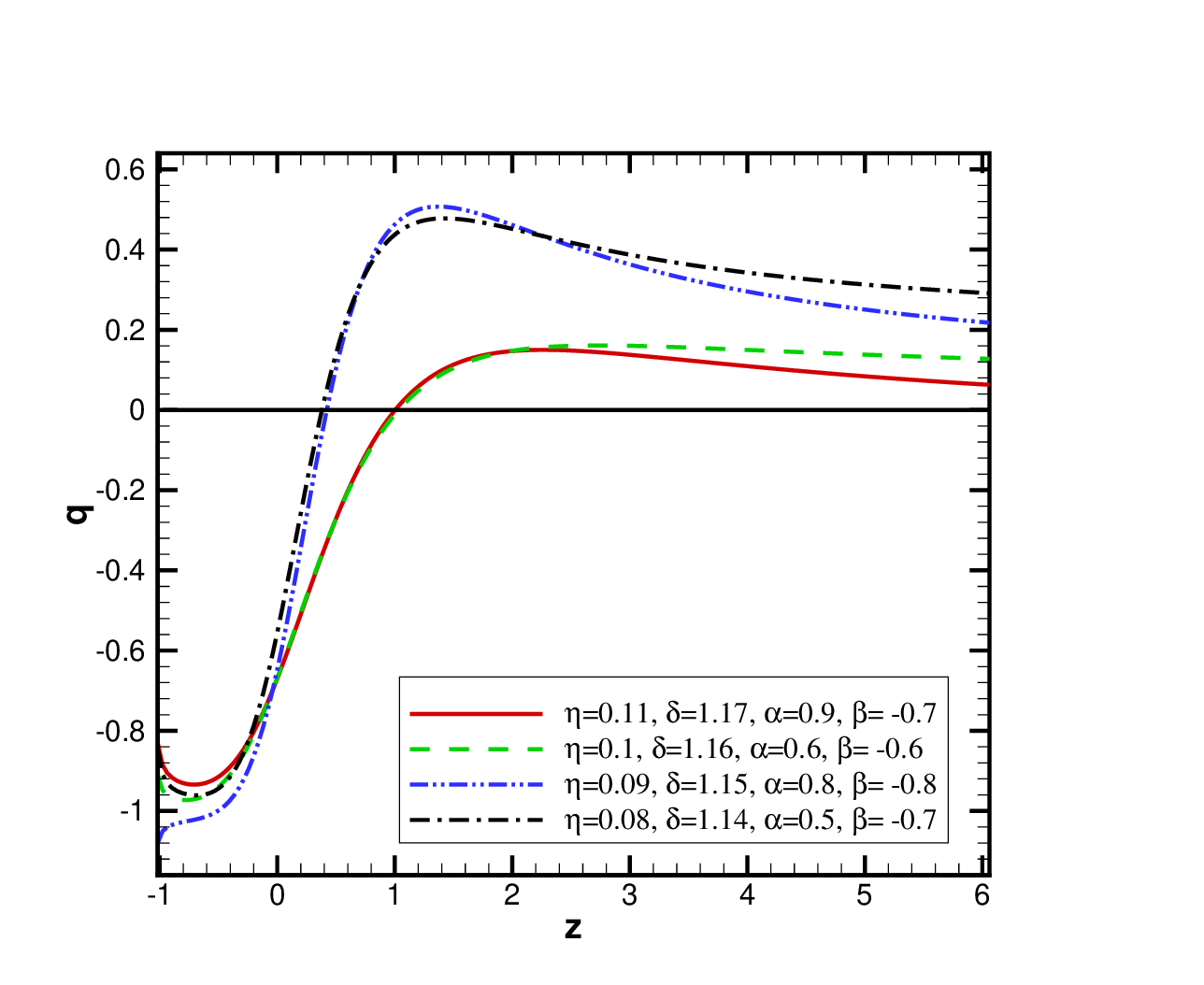}
    \caption{$ \omega_D $ and  $ q $ against $ z $
    for the interacting THDE model in Rastall theory.}\label{figwT2}
\end{center}
\end{figure}
\begin{figure}[htp]
\begin{center}
    \includegraphics[width=8cm]{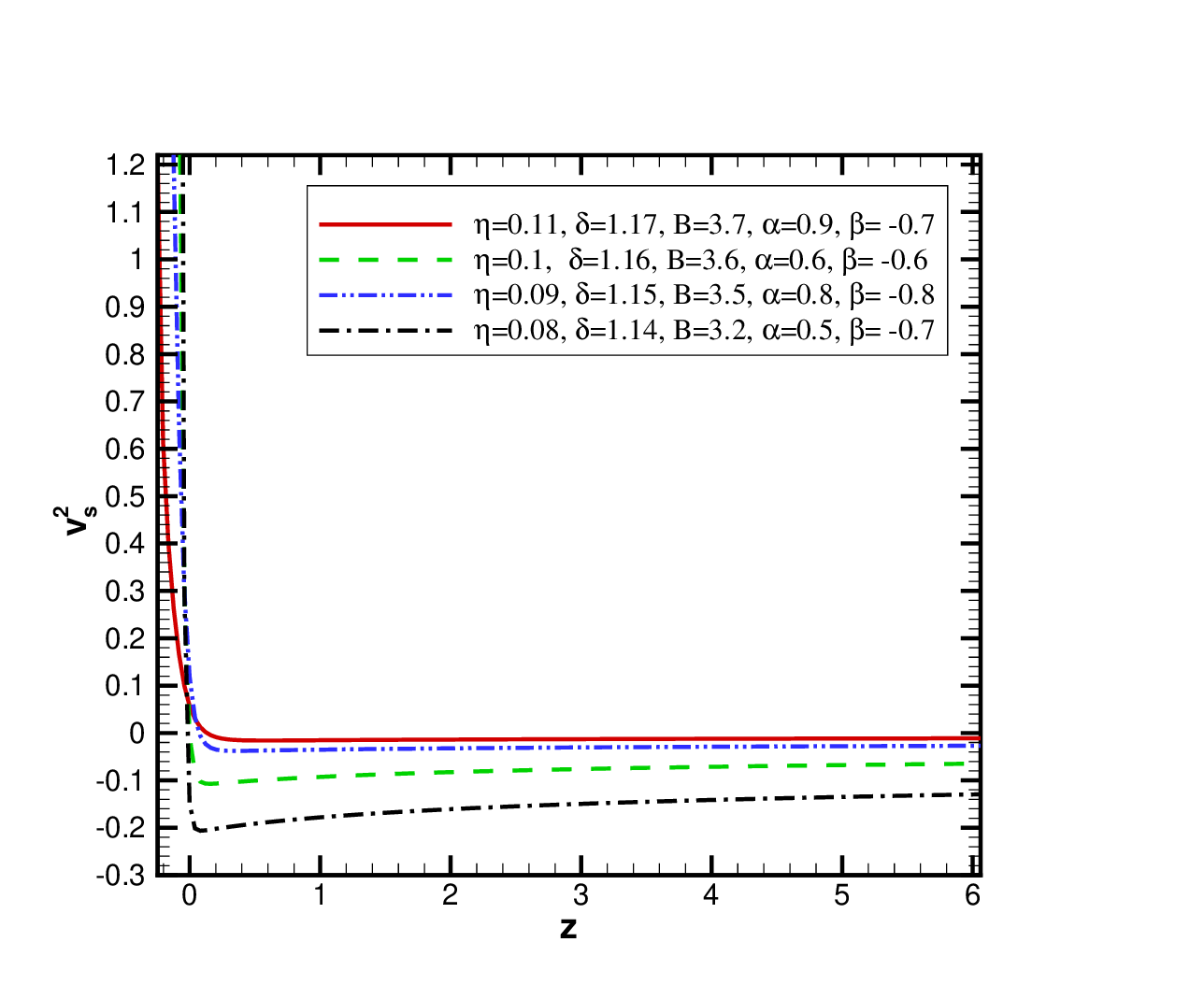}
    \caption{$ v_s^2 $ against $ z $
    for the interacting THDE model in Rastall theory.}\label{figvT2}
\end{center}
\end{figure}


\subsection*{An interacting case}
For interacting case, by using
Eqs.~(\ref{friedman3}),~(\ref{omega}),~(\ref{Qcont2}),~
(\ref{THDE}) and~(\ref{dOmegaT1}), we can find

\begin{eqnarray}\label{dOmegaeT2}
\Omega_\eta^\prime&=&\frac{4\eta-1}{(1-6\eta)(3\eta-1)}
\Big[(\alpha-3)(1-6\eta)\Omega_\eta\\&+&(4\eta-1)\Big((\alpha-3)(1-\Omega_T-6\eta)+\beta\Omega_T\Big)
\nonumber\\&+&\frac{(3\eta-1)\Omega_\eta}{\eta}((2-\delta)\Omega_T+6\eta-1)\Big]-\frac{\Omega_\eta^2}{\eta}\nonumber.
\end{eqnarray}
The behavior of the dimensionless density parameter, EoS and
deceleration parameters for interacting THDE in Rastall theory are
plotted in Figs.~\ref{figOmegaT2} and \ref{figwT2}. Figure
\ref{figOmegaT2} shows that $ \Omega_T\rightarrow 0 $ at the early
time and $ \Omega_T\rightarrow 1 $ at future ($z\rightarrow-1$),
as a desired result. It is also obvious that Universe has a
transition from the deceleration phase to the current accelerated
phase at the redshift $ z\approx 0\cdot6 $, and unlike the
non-interacting case, the interacting THDE model can produce
acceptable behavior at future, and its EoS parameter can cross the
phantom line.

Combining the time derivative of Eq.~(\ref{wT1}) with
Eqs.~(\ref{THDE}),~(\ref{dOmegaT1})~(\ref{dOmegaeT2}) and
(\ref{v_s}), one can also obtain $ v_s^2 $ for interacting THDE in
the Rastall theory. Since this expression is too long, we shall
not present it here, and only a plot of it in Fig.~\ref{figvT2} is presented, showing
that the THDE model in Rastall theory is classically stable.


\subsection{The second approach}

\noindent In this case, using
$\rho_c=\frac{3H^2(6\eta-1)}{\kappa_G(4\eta-1)}$ in rewriting the
dimensionless density parameters of THDE as

\begin{eqnarray}\label{OmegaT2}
\Omega_T=BH^{2-2\delta},
\end{eqnarray}

\noindent where $B$ is an known parameter as usual \cite{Tavayef},
and using Eq.~(\ref{OmegaD}) leading to
\begin{equation}\label{OmegaDT}
\Omega_D=\Omega_T+4\eta+2\eta\frac{\dot{H}}{H^2},
\end{equation}
and combining it with Eq.~(\ref{OmegaT2}), one reaches at
 \begin{eqnarray}\label{dOmegaT2}
 \Omega_T^{\prime}=\frac{(1-\delta)(\Omega_D-\Omega_T-4\eta)\Omega_T}{\eta}.
 \end{eqnarray}
Calculations for the EoS parameter and $q$ also yield
\begin{equation}\label{wT2}
\omega_D=\frac{-1}{3\eta\Omega_D}\Big(\Omega_D-\Omega_T-\eta\Big),
\end{equation}
and
\begin{equation}\label{qT2}
q=\frac{-1}{2\eta}(\Omega_D-\Omega_T-2\eta),
\end{equation}
respectively. It is finally useful to mention here that
Eqs.~(\ref{dOmegaT2}) ~(\ref{wT2}) and (\ref{qT2}) are valid for
both interacting and non-interacting cases.
\subsection*{Non-interacting Case}

Taking the derivative with respect to time from
Eq.~(\ref{friedman5}) and using
Eqs.~(\ref{cont1}),~(\ref{THDE}),~(\ref{OmegaDT})
and~(\ref{friedman6}), we reach at
\begin{eqnarray}\label{ddH5}
\frac{\ddot{H}}{H^3}&=&\frac{3(4\eta-1)(\Omega_D-1)}{2\eta(1-3\eta)}\nonumber\\
&-&\frac{1}{2\eta^2}((2-\delta)\Omega_T+4\eta-1)(\Omega_D-\Omega_T-4\eta).\nonumber\\
\end{eqnarray}

\noindent Now by taking the time derivative of Eq.~(\ref{OmegaDT})
and using Eqs.~(\ref{dOmegaT2}) and (\ref{ddH2}), one finds
\begin{eqnarray}\label{dOmegaD4}
\Omega_D^{\prime}=(\Omega_D-1)\Big(\frac{3(4\eta-1)}{1-3\eta}-\frac{\Omega_D-\Omega_T-4\eta}{\eta}\Big).
\end{eqnarray}
We have also plotted the behavior of $ \Omega_D(z) $, $
\omega_D(z) $ and $ q(z) $ in Figs. \ref{figOmegaT1} and
\ref{figwT1} which show that, even in the second approach, the
non-interacting THDE model can not still produce suitable behavior
at future ($z<0$).


\subsection*{An interacting Case}

The evolution of $ \Omega_D $, by combining
Eqs.~(\ref{Qcont2}),~(\ref{THDE}),~(\ref{OmegaDT})
and~(\ref{friedman6}) with the time derivative of the first
Friedmann equation~(\ref{friedman5}), is obtained as

\begin{eqnarray}\label{dOmegaD5}
\Omega_D^{\prime}&=&\frac{4\eta-1}{1-3\eta}((\alpha-3)(1-\Omega_D)+\beta\Omega_T)
\nonumber\\&-&\frac{1}{\eta}(\Omega_D-1)(\Omega_D-\Omega_T-4\eta).
\end{eqnarray}

\begin{figure}[!]
\begin{center}
    \includegraphics[width=5cm]{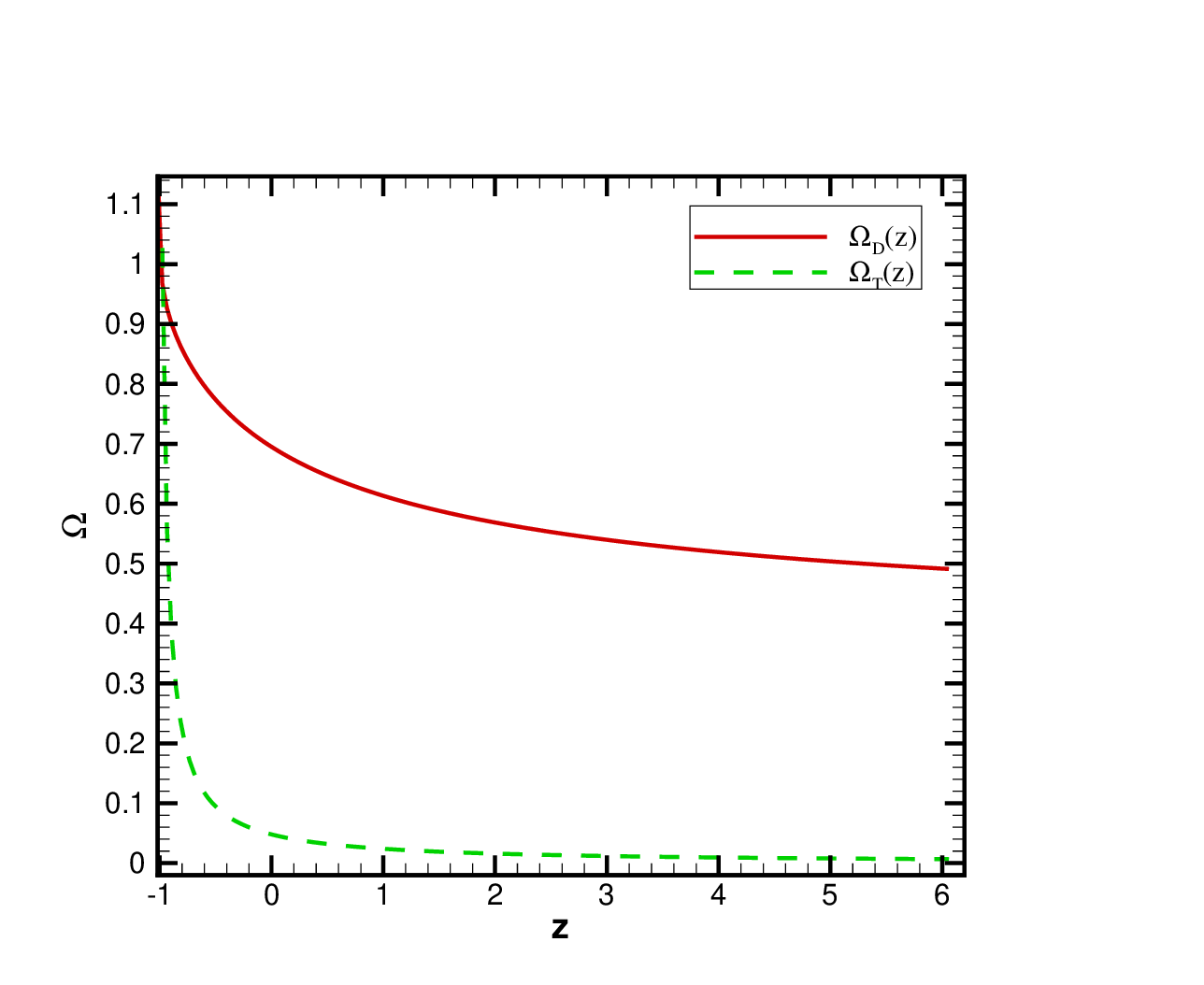}
    \caption{ $\Omega_D$ and $ \Omega_T $ versus $z$ for interacting THDE for the second approach,
     where the different parameters  $\Omega_{D}(z=0)=0\cdot73$,
     $ \eta=0.96 $, $ \delta=1.3 $, $\alpha=0.71$ and  $\beta=-0.23$
     are adopted.}\label{figOmegaDT}
\end{center}
\end{figure}

\begin{figure}[!]
\begin{center}
	\includegraphics[width=8cm]{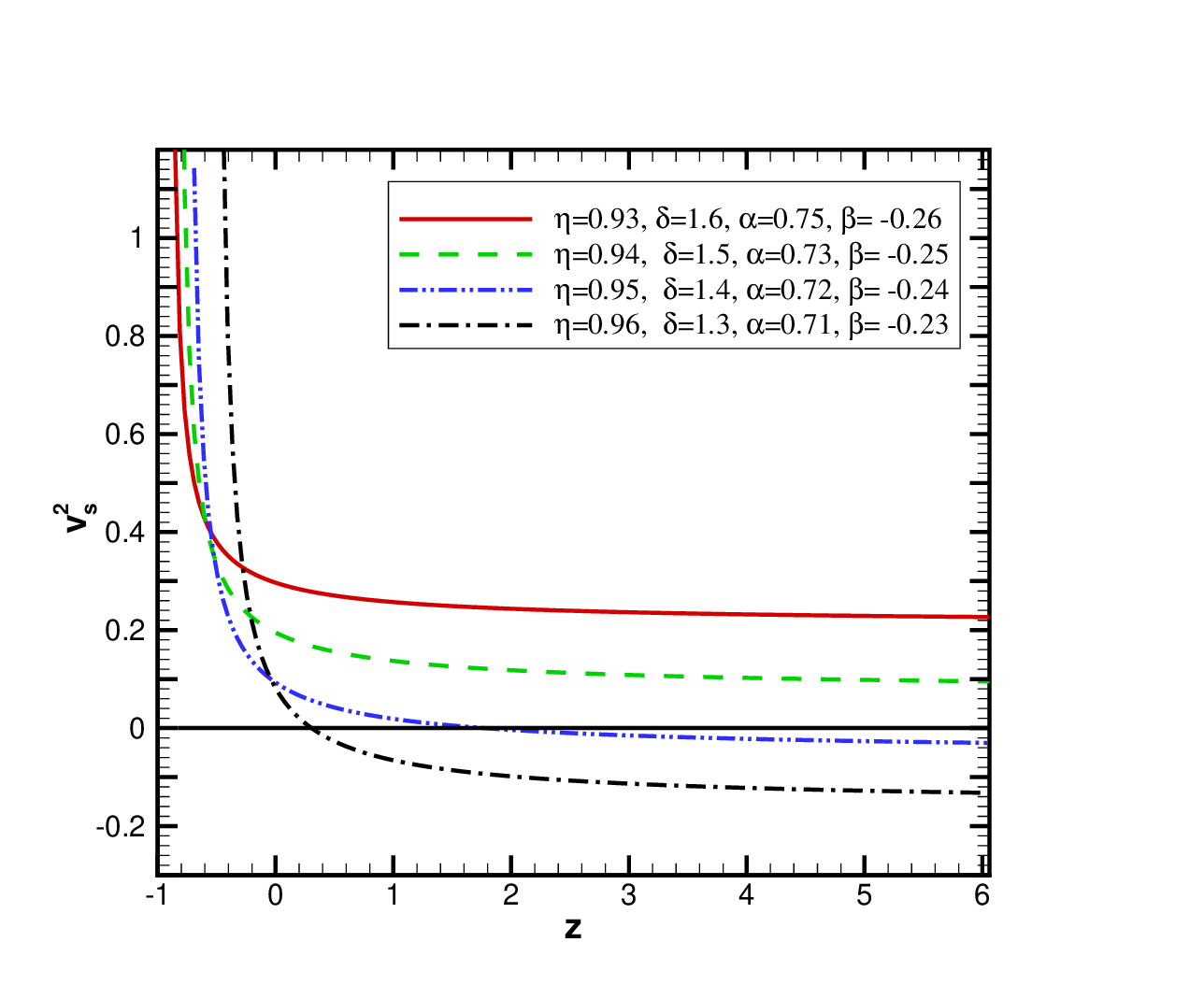}
	\caption{ $v_s^2$ and $ \Omega_T $ versus $z$ for interacting THDE model for the second approach.}\label{figvT3}
\end{center}
\end{figure}

\noindent $\omega_D $ and $ \Omega_T $ as the functions of redshift
have been plotted in Fig. \ref{figOmegaDT} indicating that $
\Omega_D\rightarrow 1 $ and $ \Omega_T\rightarrow 1 $ in future and
we have $ \Omega_T\rightarrow 0 $ and $
0\cdot1\preceq\Omega_D<0\cdot3 $ in the past. Indeed, while the
Rastall term have more contribution in $\Omega_D$ in the past, the
portion of vaccum energy ($\Omega_T$) is increased by decreasing $z$
and will get dominant at future. In Fig. \ref{figOmegaDT}, the
behavior of $ \omega_D(z) $ and $ q(z) $ are also plotted. Taking
the time derivative of relation (\ref{wT2}) and considering
Eqs.~(\ref{v_s}) and (\ref{dOmegaD4}), one can obtain $ v_s^2 $ for
interacting THDE model in the second approach. Since this expression
is too long, we have only plotted it in Fig.~\ref{figvT3} rather
than presenting it.
\begin{figure}[htp]
\begin{center}
    \includegraphics[width=8cm]{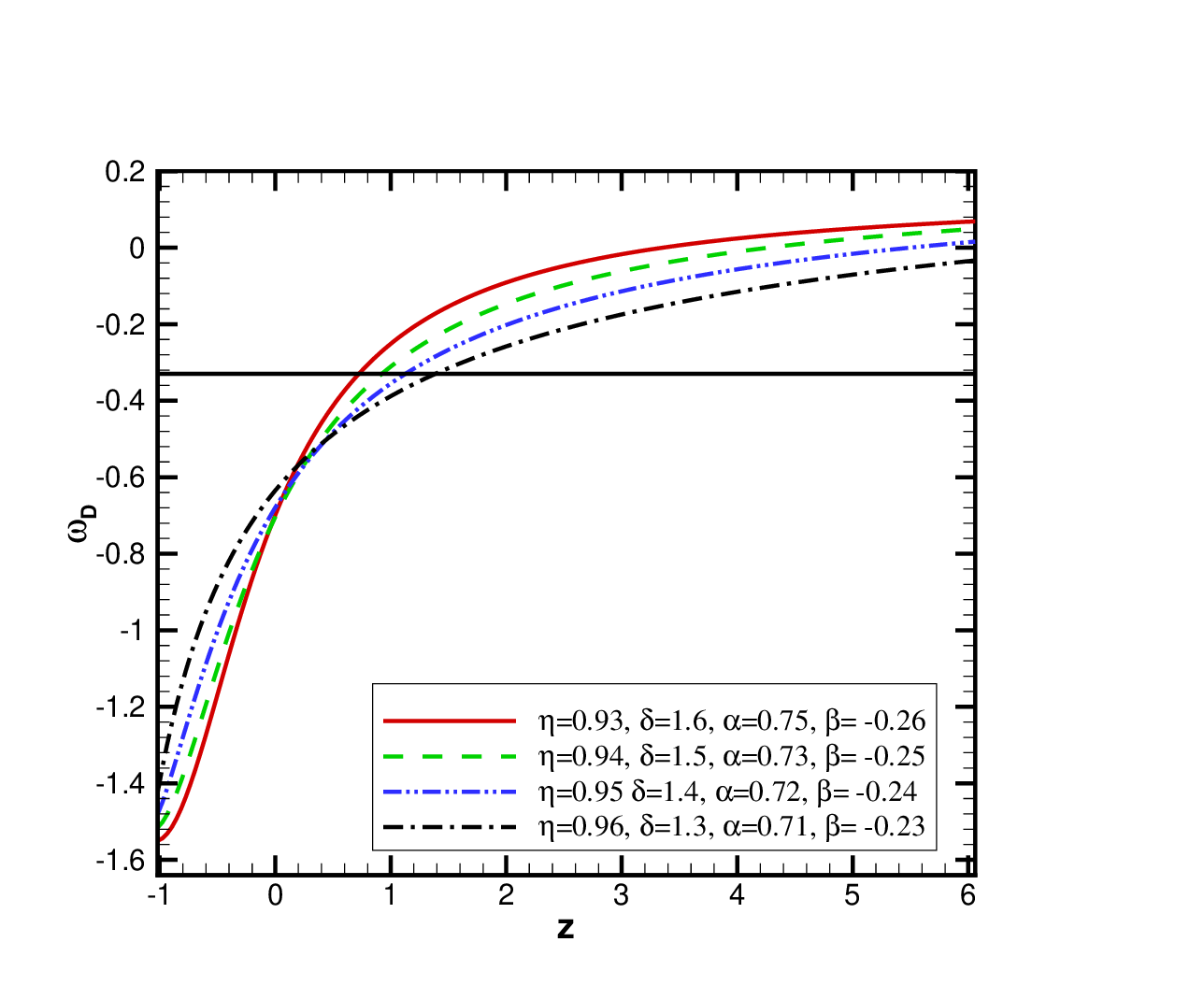}
    \includegraphics[width=7.5cm]{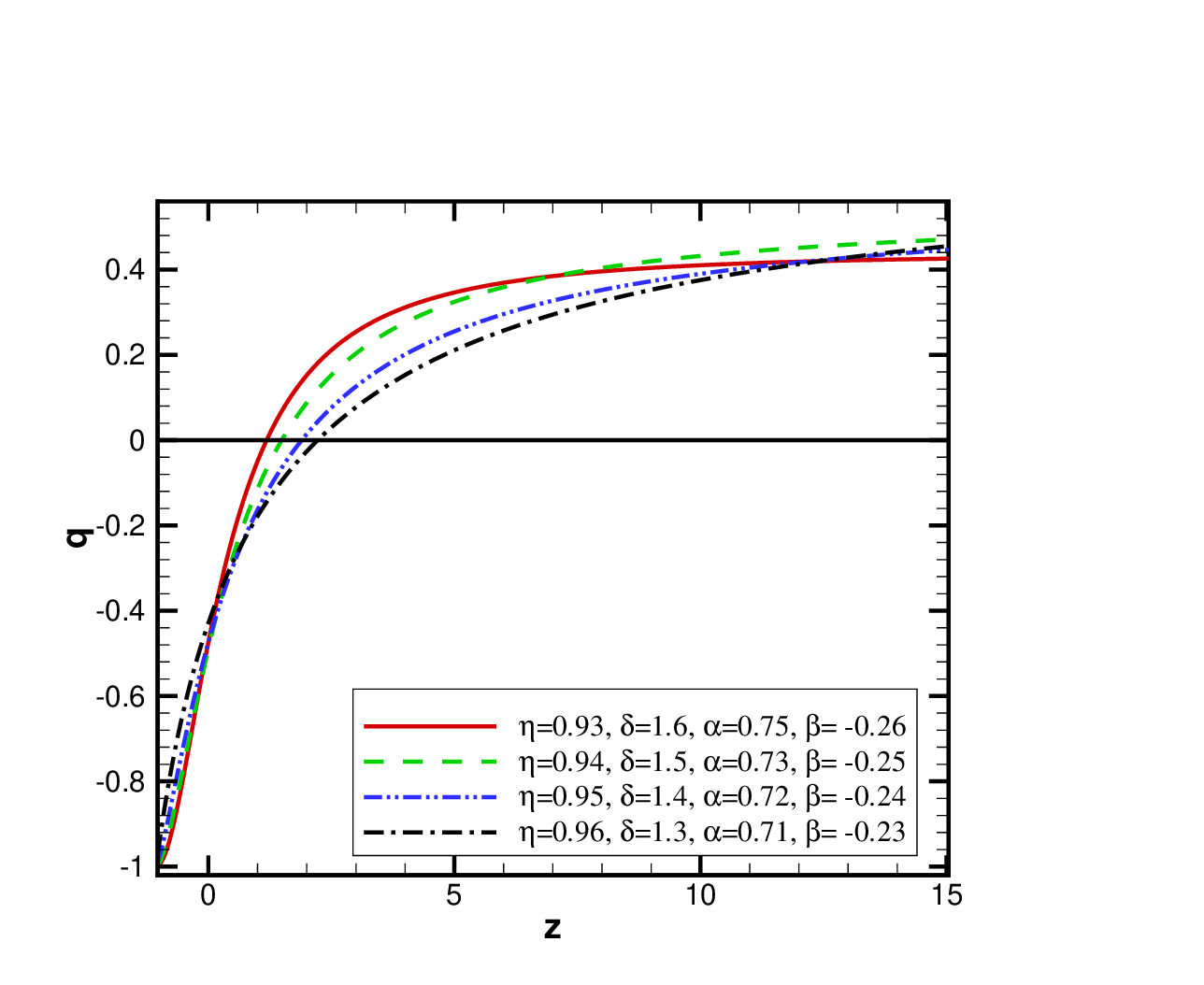}
    \caption{The evolution of the equation of state $ \omega_D $ and deceleration parameters $ q $
    with respect to the redshift $ z $ for the interacting THDE for the second approach.}\label{figwT3}
\end{center}
\end{figure}

\section{Comparision of the theoretical models with the observational Hubble parameter data}
\begin{figure}[htp]
\begin{center}
\includegraphics[width=8cm]{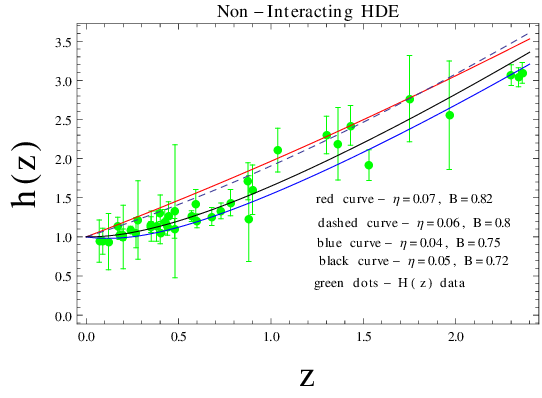}
\includegraphics[width=8cm]{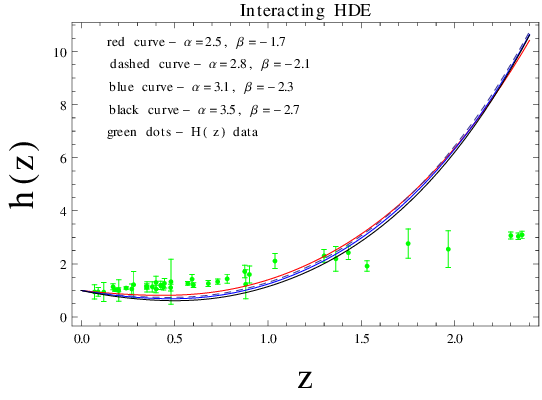}
\caption{The evolutions of the normalized Hubble parameter is shown with respect to the redshift $z$ for the non-interacting HDE model (upper panel) and the interacting HDE model (lower panel) with common approach in Rastall theory. The lower panel is for $\eta=0.11$ and $B=1.2$. The green dots correspond to the normalized Hubble parameter data consisting 41 data points \cite{hub1,hub2} with $1\sigma$ error bars.}\label{fighzhde1}
\end{center}
\end{figure}

\begin{figure}[htp]
\begin{center}
\includegraphics[width=8cm]{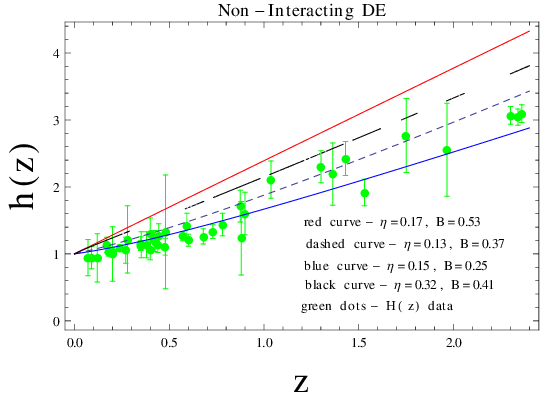}
\includegraphics[width=8cm]{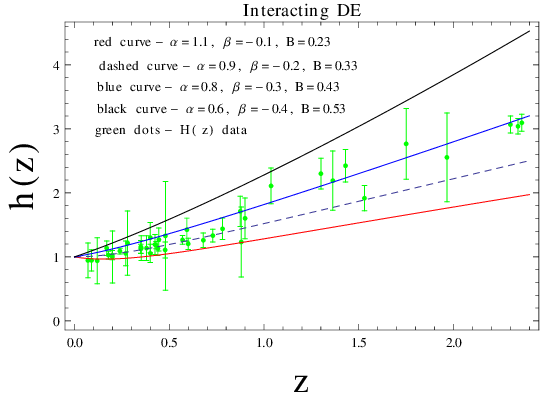}
\caption{The evolutions of the normalized Hubble parameter is shown with respect to the redshift $z$ for the non-interacting DE model (upper panel) and the interacting DE model (lower panel) with new approach in Rastall theory. The lower panel is for $\eta=0.1$.}\label{fighzhde2}
\end{center}
\end{figure}

\begin{figure}[htp]
\begin{center}
\includegraphics[width=8cm]{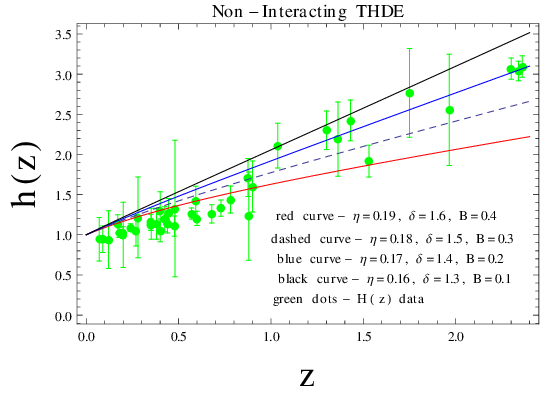}
\includegraphics[width=8cm]{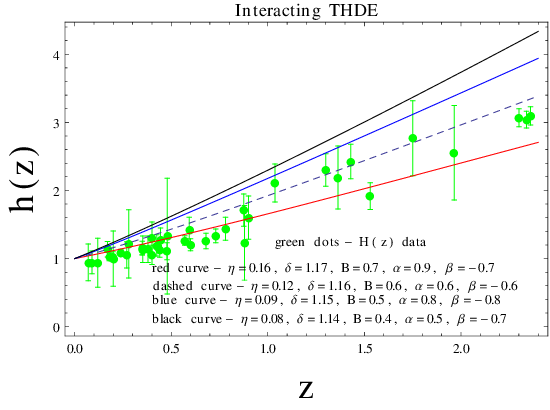}
\caption{The evolutions of the normalized Hubble parameter is shown with respect to the redshift $z$ for the non-interacting THDE model (upper panel) and the interacting THDE model (lower panel) with common approach in Rastall theory.}\label{fighzthde1}
\end{center}
\end{figure}

\begin{figure}[htp]
\begin{center}
\includegraphics[width=8cm]{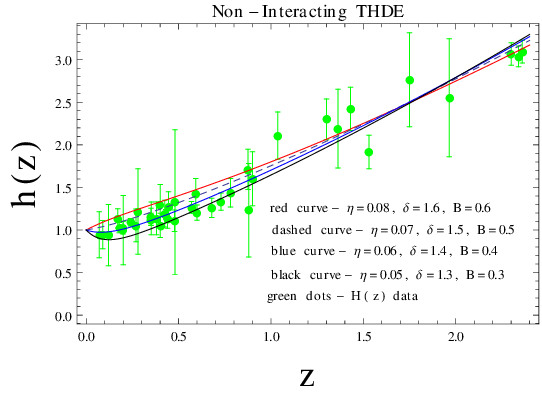}
\includegraphics[width=8cm]{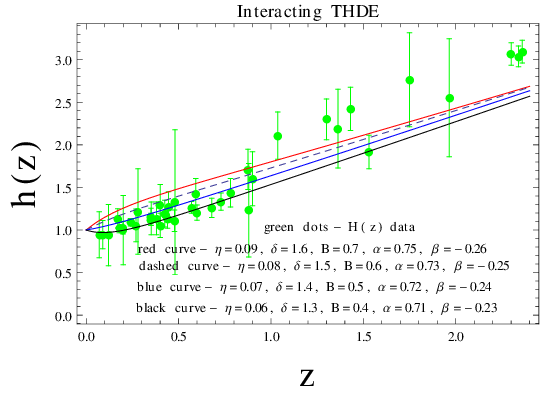}
\caption{The evolutions of the normalized Hubble parameter is shown with respect to the redshift $z$ for the non-interacting THDE model (upper panel) and the interacting THDE model (lower panel) with new approach in Rastall theory.}\label{fighzthde2}
\end{center}
\end{figure}

In Figs. \ref{fighzhde1}-\ref{fighzthde2}, we have shown the evolution of normalized Hubble parameter $h(z)=\frac{H(z)}{H_0}$ for our models and compared it with that of the data points for normalized Hubble parameter with $1\sigma$ error bars which have been obtained from the compilation of 41 points of $H(z)$ measumements \cite{hub1,hub2} using the present value of $H(z)$ from \cite{r16}. The corresponding error in $h(z)$ is given as \cite{zt4}
\begin{equation}
\sigma_{h}=h\sqrt{\frac{\sigma^{2}_{H_0}}{H^{2}_0} + \frac{\sigma^{2}_{H}}{H^{2}}}
\end{equation}
where, $\sigma^{2}_{H}$, $\sigma^{2}_{H_0}$ are errors
in $H$ and $H_0$ respectively.\\

 We found that for most of the cases, our models quite fit well with the latest $H(z)$ dataset at low redshifts. Furthermore, we also checked that the nature of the evolution of $h(z)$ is hardly affected by a small change in the values of the model parameters. Therefore, we conclude that our models are consistent with the $H(z)$ data in the comfortable region of the parameter space.
\section{Summary}

We found out that, the same as the original HDE in standard
cosmology \cite{HDE3}, the density dimensionless parameter of HDE in
Rastall framework is constant in common approach. The behavior of
deceleration and EoS parameters is suitable in both interacting and
non-interacting regimes. In the second approach, while the sum of
HDE and Rastall parameter plays the role of dark energy, unlike the
common approach, the density dimensionless parameter is not constant
during the cosmic evolution. In this manner and in the presence of
interaction, all cosmological parameters show acceptable behavior
with the same values of the coupling constants including $\eta$,
$\alpha$ and $\beta$, an outcome which may not be obtained in the
absence of the assumed mutual interaction between vacuum energy and
dark matter.\\
Our study implies the fact that the density dimensionless
parameter of THDE ($\Omega_T$) is not always constant during the cosmic
evolution, and the cosmic parameters show proper behavior by themselves
only whenever the mutual interaction is present between the vacuum
energy and dark matter. The same conclusion is also valid in the
second approach. It is finally useful to mention that only the
non-interacting HDE in common approach is always classically
unstable. We also found that the models are consistent with the latest $H(z)$ data at low redshifts for a wide range of the values for the models parameters.

\end{document}